\documentclass[%
preprint,
amsmath,amssymb,
prb,
]{revtex4}

\usepackage{graphicx}
\usepackage{dcolumn}
\usepackage{bm}

\begin{document}


\title{Lattice QCD study of four-quark components of the isosinglet scalar mesons: 
Significance of disconnected diagrams
}

\author{Masayuki Wakayama$^1$}%
\author{Teiji Kunihiro$^2$}
\author{Shin Muroya$^3$}%
\author{Atsushi Nakamura$^4$}%
\author{Chiho Nonaka$^{1, 5}$}%
\author{Motoo Sekiguchi$^6$}%
\author{Hiroaki Wada$^7$}%

\affiliation{%
$^{1}$Department of Physics, Nagoya University, Nagoya 464-8602, Japan\\
$^{2}$Department of Physics, Kyoto University, Kyoto 606-8502, Japan\\
$^{3}$Faculty of Comprehensive Management, Matsumoto University, Matsumoto 390-1295, Japan\\
$^{4}$RCNP, Osaka University, Osaka 567-0047, Japan\\
$^{5}$Kobayashi Maskawa Institute, Nagoya University, Nagoya 464-8602, Japan\\
$^{6}$School of Science and Engineering, Kokushikan University, Tokyo 154-8515, Japan\\
$^{7}$Faculty of Political Science and Economics, Kokushikan University, Tokyo 154-8515, Japan
}%

\collaboration{SCALAR Collaboration}



\date{\today}

\begin{abstract}
We study the possible significance of four-quark states in the iso-singlet scalar mesons ($J^{PC}=0^{++}$, $I=0$)  
by performing two-flavor full lattice QCD simulations on an $8^3 \times 16$ lattice using the improved gauge action 
and the clover-improved Wilson quark action. 
In particular, 
we evaluate the propagators of molecular and tetra-quark operators 
together with singly disconnected diagrams. 
In the computation of the singly disconnected diagrams 
we employ the $Z_2$-noise method with the truncated eigenmode approach. 
We show that 
the quark loops given by the disconnected diagrams 
play an essential role in propagators of tetraquark and molecular operators. 


\end{abstract}

\maketitle


\def\ka#1{\left( #1 \right)}
\def\kak#1{\left\{ #1 \right\}}
\def\kakk#1{\left[ #1 \right]}
\def\ex#1{\mathrm{e}^{#1}}
\def\Tr{\mathrm{Tr}}
\def\bpsi{\bar{\psi}}
\def\mbf#1{\mathbf{#1}}
\def\non{\nonumber}
\def\qprop{{\small{\ka{W^{-1}}}}}


\section{Introduction}
The approximate chiral symmetry and its 
spontaneous breaking in QCD are indispensable basic ingredients for understanding 
the low-energy phenomena of hadrons. 
The pions should be the remnants of the Nambu-Goldstone (NG) bosons 
associated with the spontaneous breaking of 
chiral SU(2)$\otimes$SU(2) symmetry with 
$\langle \bar{u}u+\bar{d}d\rangle/\sqrt{2}$ being the order parameter; their small masses come from 
the tiny current quark masses of $u$ and $d$ quarks. The other would-be
NG boson is the $\eta$, which is massive even in the chiral limit where the
quarks are massless due to the axial anomaly in QCD.
In the linear representation of SU(2)$\otimes$SU(2), 
the four scalar bosons appear, one of which is traditionally
called the $\sigma$ meson. The scalar bosons are the amplitude fluctuations of 
the chiral order parameter, while the NG bosons are the phase fluctuations.
In view of the success of the nonlinear realization 
of the chiral symmetry in describing the low-energy hadron phenomena\cite{Weinberg2}, 
the curvature of the effective potential might be large, and accordingly 
the $\sigma$ might appear only as a high-lying state coupled with other states.
Nevertheless, the picture given by the linear representation where the $\sigma$ exists 
as a basic ingredient should become relevant around the (pseudo-) critical
region of the chiral transition, which is found to be a crossover with a transitional
region in the lattice QCD.

Interestingly enough,
recent experiments and precise and systematic analyses 
of the $\pi$-$\pi$ scattering respecting the 
crossing symmetry as well as the chiral symmetry have revealed the existence of 
the low-mass scalar meson with a mass from 400 to 700\, MeV\cite{PDG14}.
The  physical content and the mechanism for 
realizing such a low-lying state in the $J^{PC}=0^{++}$ state have 
prompted much debate. 
One of the most popular ideas is that all the low-lying scalar states
can be realized as tetraquark states, i.e., diquark-antidiquark states, 
as first advocated by Jaffe\cite{Jaffe1}. 
On the other hand, the appearance of the $\sigma$ in the 
$\pi$-$\pi$ scattering may simply suggest that the meson is a
$\pi$-$\pi$ resonance state with the pion maintaining its identity;
if the pions were heavy, the resonance state may turn into a molecular state of the heavy pion.
Note that such four-quark states, irrespective 
of whether they are molecular states or tetraquark states, 
are more likely to exist in the heavy-quark sectors:
such exotic states include $X(3872)$, $Y(4260)$, $Z(4430)$, $Z_b(10610)$, and $Z_b(10650)$
\cite{X3872,Y4260,Z4430,Zb10610_Zb10650}. 
It would be interesting to see how the four-quark states or the components of a hadron 
change as the quark masses are changed. 

In the present work, 
we explore the possible significance of 
the four-quark components in the iso-singlet scalar mesons 
by performing two-flavor full lattice QCD simulations. 
Many quenched lattice simulations have been carried out for the isosinglet scalar mesons 
\cite{Jaffe,Suganuma:2007uv,Mathur:2006bs,Loan:2008sd,Sasa09}. 
The first full QCD calculation of the $\sigma$ meson
was performed by the SCALAR Collaboration\cite{Scalar04}, 
where the $(\bar q q)$ interpolation field was used 
and a disconnected diagram, 
i.e., a quark-loop diagram in the normal language of the quantum field theory, 
was evaluated using the $Z_2$-noise method 
with the truncated eigenmode approach\cite{McNeile:2000xx,Struckmann:2000bt,Neff:2001zr}. 
It was found that the inclusion of the disconnected diagram 
is indispensable for obtaining a clear signal showing the existence of 
the low-lying scalar meson. 
They also showed a significant quark mass dependence of the 
clearness and the resultant mass of the $\sigma$.   
There have been many subsequent studies of the scalar mesons including 
the $\kappa$ based on lattice simulations of 
full QCD\cite{UKQCD06_1,Scalar07,UKQCD06_2,BGR12}.
The possible four-quark nature of the isononsinglet scalar mesons 
has also been examined on a lattice\cite{ETM13}. 
This work was continued in Ref. \cite{a0} where 
the technical aspects for computation of the tetraquark candidate
 $a_0(980)$ including disconnected diagrams were discussed 
 and the preliminary results were shown. 
More recently, Prelovsek {\it et al.}~\cite{Sasa10} explored the possibility
that the $\sigma$ meson is well described as a four-quark state, i.e.,
a molecular or tetraquark state, without taking into account
the disconnected diagrams, which may unfortunately make the physical
significance of their result obscure in view of the essential significance 
of the disconnected diagram observed in Ref.~\cite{Scalar04}.
We show that 
the quark loops given by the disconnected diagrams 
play an essential role in making the four-quark states exist. 
We perform simulations 
both with and without disconnected diagrams and compare them.
Although the quark masses used in the present work 
are admittedly not small, and hence it may not be straightforward to extract
direct implications regarding the nature of the $\sigma$, 
our work may be an important 
milestone to understand  the role of the four-quark states possibly changing 
from light to heavy quark sectors. 

In the present work, we prepare two types of interpolation operators for the creation of four-quark states: 
a molecular operator $(\bar q q)(\bar q q)$ with $(\bar q q)$ 
and a tetraquark operator $(\bar q \bar q q q)$ 
composed of a diquark $(q q)$ and an antidiquark $(\bar q \bar q)$ being color singlet. 
There are many other operators with the same quantum number as the $\sigma$, which include
$(\bar q q)$, $(\bar q q)(\bar q q)$, $(\bar q \bar q q q)$,
the glueballs $(gg)$, the hybrids $(\bar q q g)$, and their excited modes.
It would certainly be desirable to include all the operators for a precise calculation. 
In the present work, however, we do not include these operators as interpolation operators. 
Note that our calculation is a full QCD calculation and 
hence all the states that couple to the $\sigma$ 
should, in principle, be created in the intermediate states, 
provided that the prepared interpolation operators well coupled with these states. 
Moreover, it has been reported \cite{Chen06} that 
the scalar glueball is heavy with a mass of approximately $1500$\, MeV 
and will be decoupled from the low-lying $\sigma$. 
Therefore, the neglect of the glueball state as well as hybrids including 
glueball states should be valid for the description of the $\sigma$.
Needless to say, the numerical cost will become huge 
for full QCD calculations incorporating all the above interpolation operators. 
This numerical cost is especially huge when we include the disconnected diagrams. 

The present article is organized as follows. 
We begin in Sec. II by showing the formulation of the four-quark propagators. 
In Sec. III we give the numerical results of our simulations 
and discuss the significance of the disconnected diagrams 
in the four-quark propagators, effective masses, and the isosinglet scalar mesons. 
We end in Sec. IV with our conclusions.

\section{Four-quark states in the isosinglet scalar mesons}
We investigate the possible significance  of the four-quark states in the isosinglet scalar mesons 
by performing the two-flavor full lattice QCD simulations. 
In particular, focusing on the ingredients in the four-quark states that might consist  of 
the molecular state $(\bar q q)(\bar q q)$ and/or the tetraquark state $(\bar q \bar q q q)$, 
we prepare two types of operators for four-quark states. 

The molecular interpolation operators are defined as 
\begin{widetext}
\begin{eqnarray}
{\cal{O}}^{\rm{molec}}(t) &=& \frac{1}{\sqrt{3}}
                                        \kakk{{\cal{O}}^{\pi ^{+}}(t){\cal{O}}^{\pi^{-}}(t)
                                                -{\cal{O}}^{\pi ^{0}}(t){\cal{O}}^{\pi^{0}}(t)
\label{eq-molec}                                               +{\cal{O}}^{\pi ^{-}}(t){\cal{O}}^{\pi^{+}}(t) } \ , 
\end{eqnarray}
\end{widetext}
where ${\cal{O}}^{\pi^+}(t)$, ${\cal{O}}^{\pi^-}(t)$, and ${\cal{O}}^{\pi^0}(t)$ 
are the $\pi$ meson operators made up of two quarks. 
They are given by 
\begin{eqnarray}
{\cal{O}}^{\pi ^{+}}(t) &=& -\sum_{\mbf{x}\, a} 
                                                 \bar d^{a}(t,\mbf{x})\gamma _5 u^{a}(t,\mbf{x}) \ , \non \\
{\cal{O}}^{\pi ^{-}}(t) &=& \sum_{\mbf{x}\, a} 
                                               \bar u^{a}(t,\mbf{x})\gamma _5 d^{a}(t,\mbf{x}) \ , \non \\
{\cal{O}}^{\pi ^{0}}(t) &=& \frac{1}{\sqrt 2} \sum_{\mbf{x}\, a}
                                                \left[ \bar u^{a}(t,\mbf{x})\gamma _5 u^{a}(t,\mbf{x})  
                                                      - \bar d^{a}(t,\mbf{x})\gamma _5 d^{a}(t,\mbf{x})\right] \ , 
\end{eqnarray}
where $a$ is the index of the color. 

The tetraquark interpolation operators are given by 
\begin{eqnarray}
{\cal{O}}^{\rm{tetra}}(t) &=& \sum_{a} [u d ]^a (t) [\bar u \bar d ]^a (t) \ , 
\end{eqnarray}
where $[u d ]^a (t)$ and $[\bar u \bar d ]^a (t)$ are diquark and antidiquark operators, respectively, written as 
\begin{widetext}
 \begin{eqnarray}
[u d ]^a (t) &=&  \frac{1}{2} \sum_{\mbf{x}\, b,c}  
                          \epsilon^{abc} \left[ u^{T b}(t,\mbf{x}) \  C\gamma_5 \ d^{c}(t,\mbf{x}) 
                                                       - d^{T b}(t,\mbf{x}) \ C\gamma_5 \ u^{c}(t,\mbf{x}) \right] \ , \ \ \ \ \non \\
\ [ \bar u \bar d ]^a (t) &=&  \frac{1}{2} \sum_{\mbf{x}\, b,c}  
                          \epsilon^{abc} \left[ \bar u^{b}(t,\mbf{x}) \  C\gamma_5 \ \bar d^{T c}(t,\mbf{x})  
                                                       - \bar d^{b}(t,\mbf{x}) \  C\gamma_5 \ \bar u^{T c}(t,\mbf{x}) \right] \ , \ \ \ \ \ \ \ \ \ \ 
\end{eqnarray}
\end{widetext}
with the charge conjugation matrix $C$. 

For the interpolation operators of the molecule and tetraquark there are other possible candidates. 
For example, in Ref.~\cite{Sasa10}, 
vector- and axial-vector-type operators as well as pseudoscalar-type operators 
were used for the molecule. 
For the tetraquark, 
in addition to the (anti)pseudoscalar diquark operators, the (anti)scalar diquark operators are also employed. 
The choice of the operators for the molecule and tetraquark  
is motivated by the fact that pseudoscalar mesons are the lightest mesons 
and diquarks with $C\gamma_5$ are the lightest diquarks \cite{Alexandrou:2006cq,Jaffe:2004ph, Wagner:2011fs}. 

The propagator $G^{i}(t)$ for the four-quark operators is written as 
\begin{eqnarray}
G^{i}(t) 
 &=& \left \langle {\cal{O}}^{i}(t) {\cal{O}}^{i \, \dag}(0) \right \rangle  \ , \ i = {\rm molec  \ or \  tetra}, 
\label{Gij}
\end{eqnarray}
where ${\cal{O}}^{i}$ is the molecular or tetraquark interpolation operator. 

We show the diagrams for the elements of the propagator $G^{i}(t)$: 
the molecule  $G^{\rm{molec}}(t)$ and  
the tetraquark $G^{\rm{tetra}}(t)$. 
Through the functional integral of Eq.~(\ref{Gij}) with the quark fields,  
the propagator of the molecular operator  $G^{\rm{molec}}(t)$ is written as 
\begin{widetext}
\begin{eqnarray}
G^{\rm{molec}}(t) \!\!&=&\!\! 2\kakk{ D(t) + \frac{1}{2} C(t) - 3 A(t) + \frac{3}{2} V(t) } \ ,  
\label{molec-molec}
\end{eqnarray}
\end{widetext}
\begin{figure}[h]
\begin{center}
\includegraphics[scale=0.24]{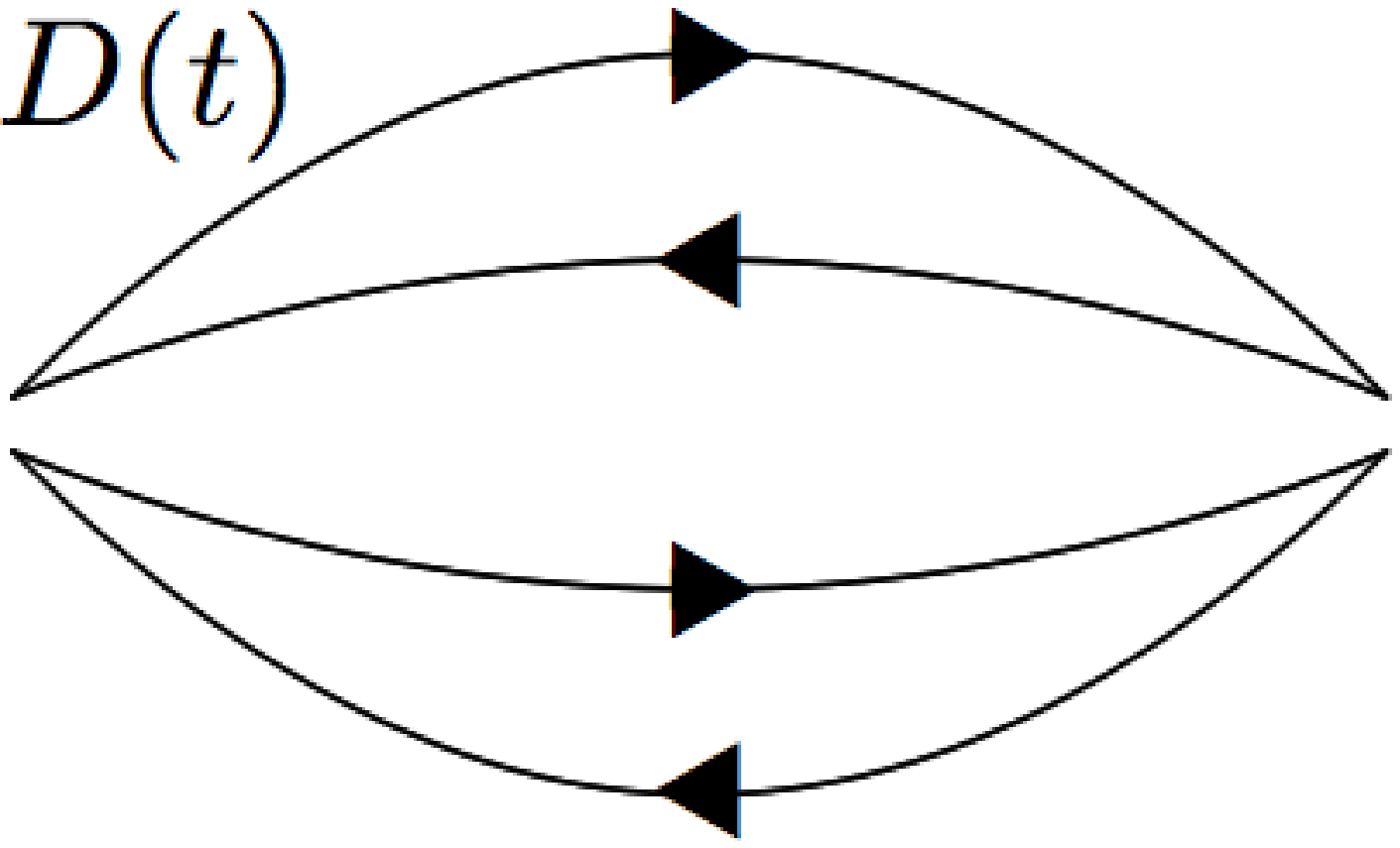} \ \ \ 
\includegraphics[scale=0.24]{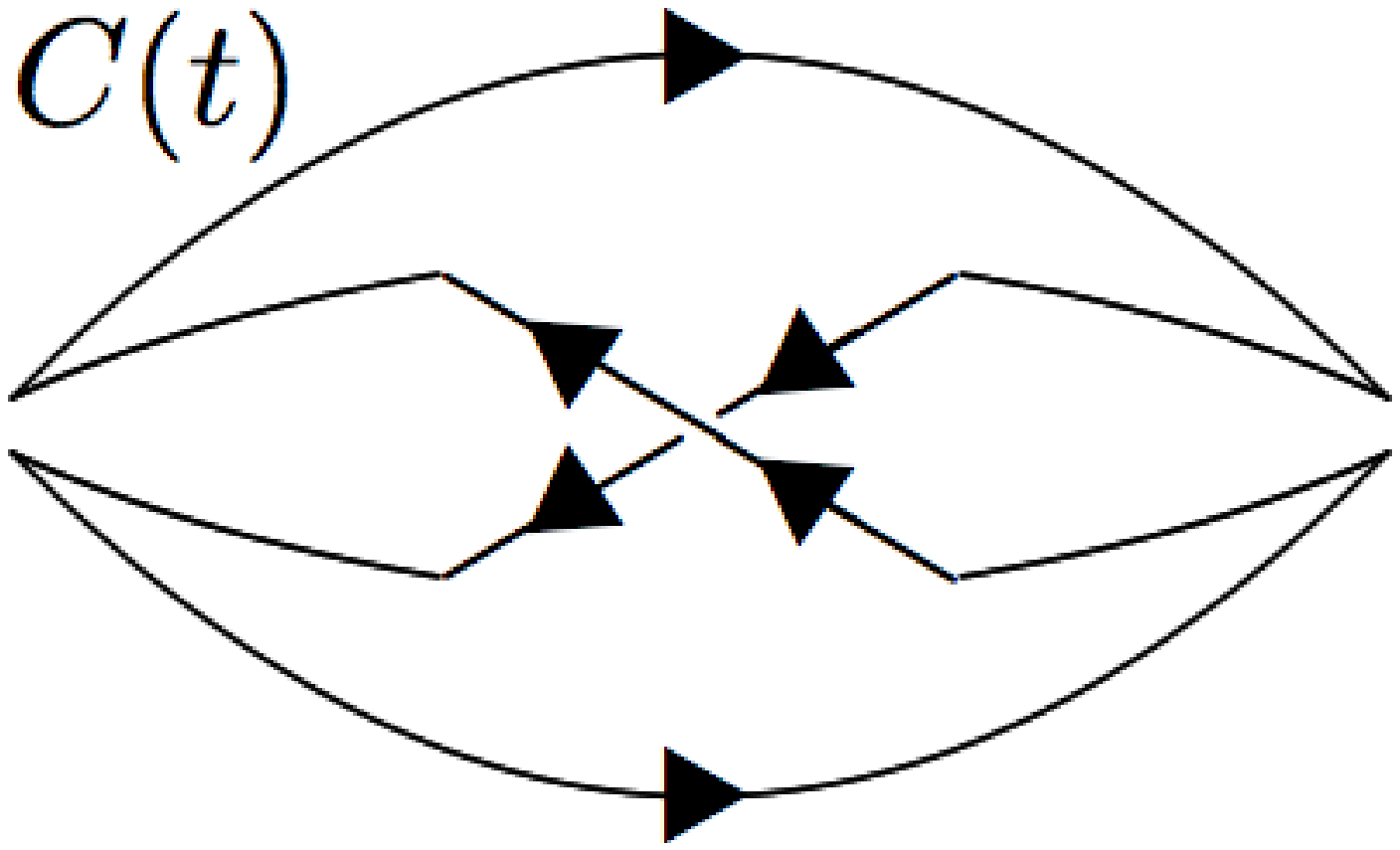} \ \ \ 
\includegraphics[scale=0.24]{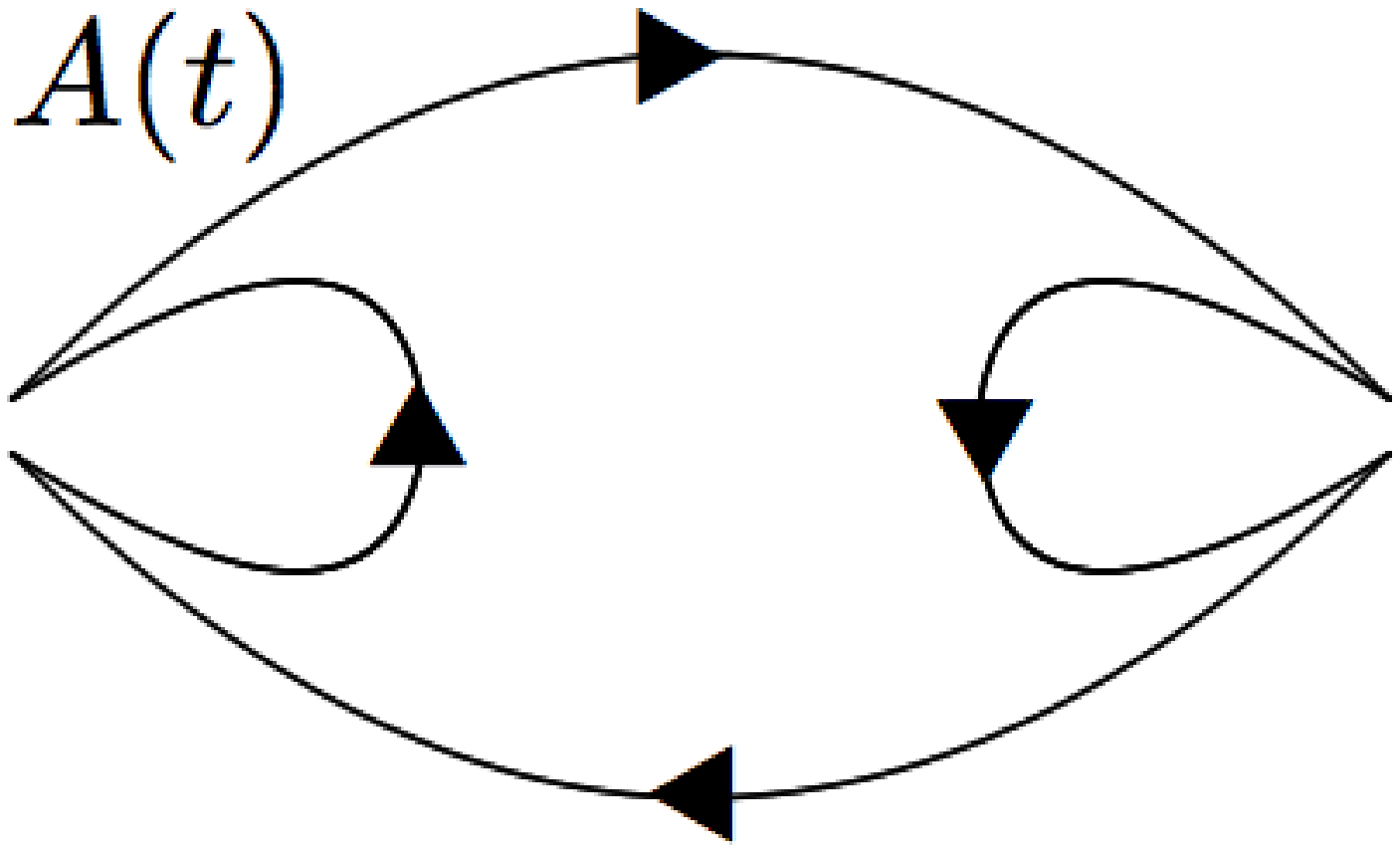} \ \ \ 
\includegraphics[scale=0.24]{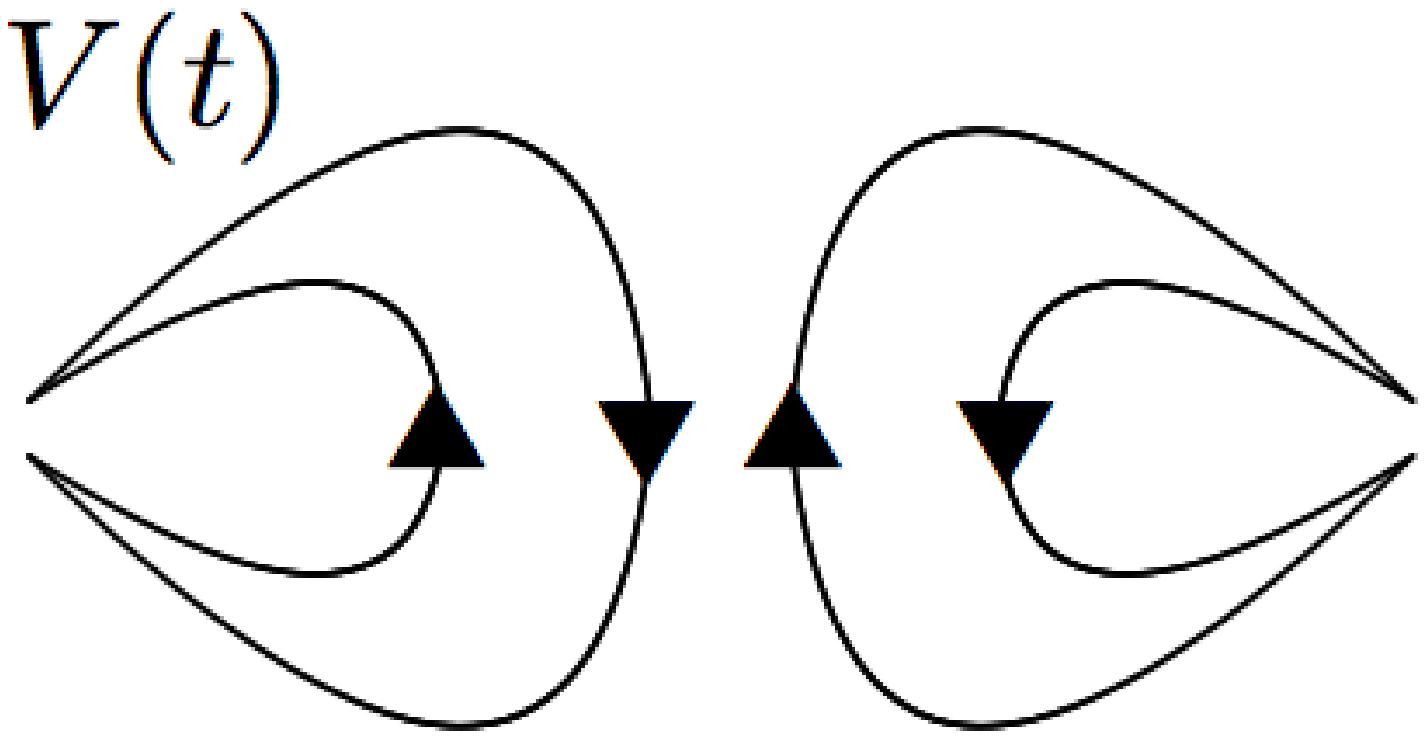}
\caption{
The diagrams for the propagator of the molecular operator  $G^{\rm{molec}}(t)$.}
\label{DCAG}
\end{center}
\end{figure}
where $D(t)$, $C(t)$, $A(t)$, and $V(t)$ correspond to 
direct, crossed, single annihilation (singly disconnected), 
and vacuum (doubly disconnected) diagrams, respectively  (Fig.~\ref{DCAG}). 
The detailed expression for each diagram is given in the Appendix. 
The tetraquark propagator is given by 
\begin{widetext}
\begin{eqnarray}
G^{\rm{tetra}}(t) \!\!&=&\!\! 2\ka{D^{\prime}_{1}(t)+D^{\prime}_{2}(t)}
 - 2\ka{A^{\prime}_{1}(t)+A^{\prime}_{2}(t)+A^{\prime}_{3}(t)+A^{\prime}_{4}(t)} 
 \ \ \ \  \nonumber \\
& & \ \ \ \ \ \ \ \ \ \ \ \ \ \ \ \ \ \ \ \ \ \ \ \ \  
 + \ka{V^{\prime}_{1}(t)+V^{\prime}_{2}(t)+V^{\prime}_{3}(t)+V^{\prime}_{4}(t)} \ , 
 \label{tetra-tetra}
\end{eqnarray}
\end{widetext}
\begin{figure}[h]
\begin{center}
\includegraphics[scale=0.24]{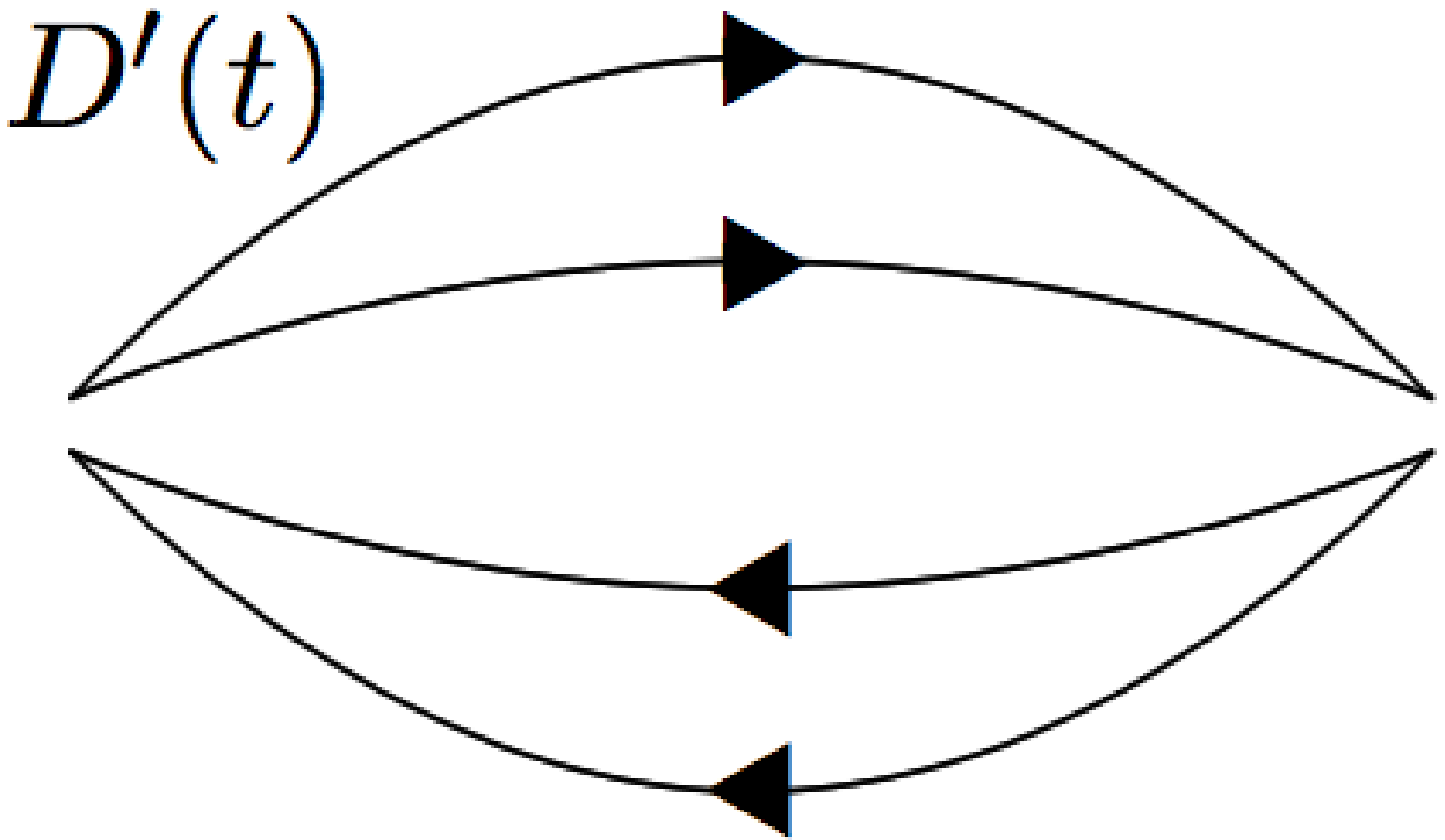} \ \ \ 
\includegraphics[scale=0.24]{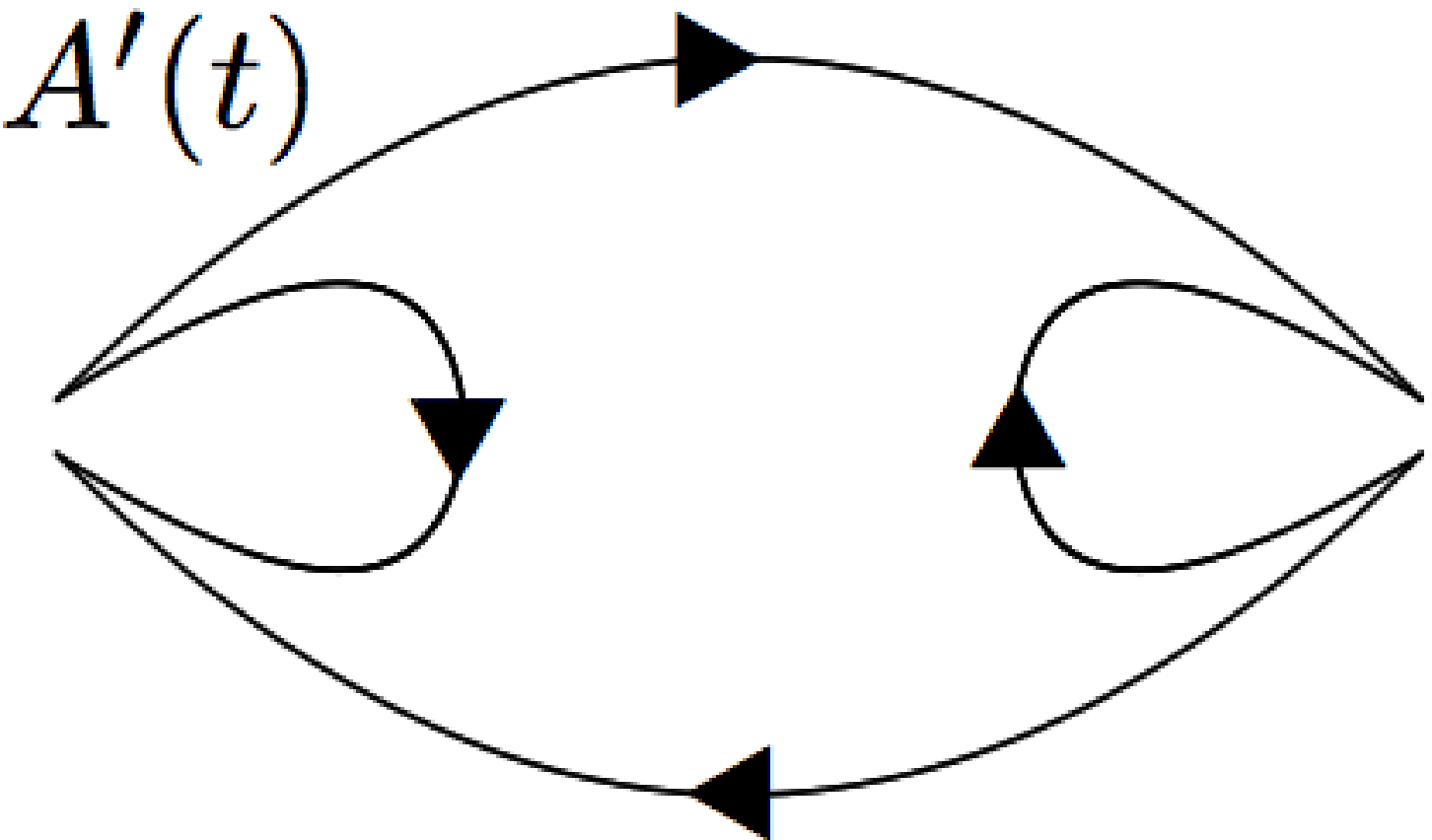} \ \ \ 
\includegraphics[scale=0.24]{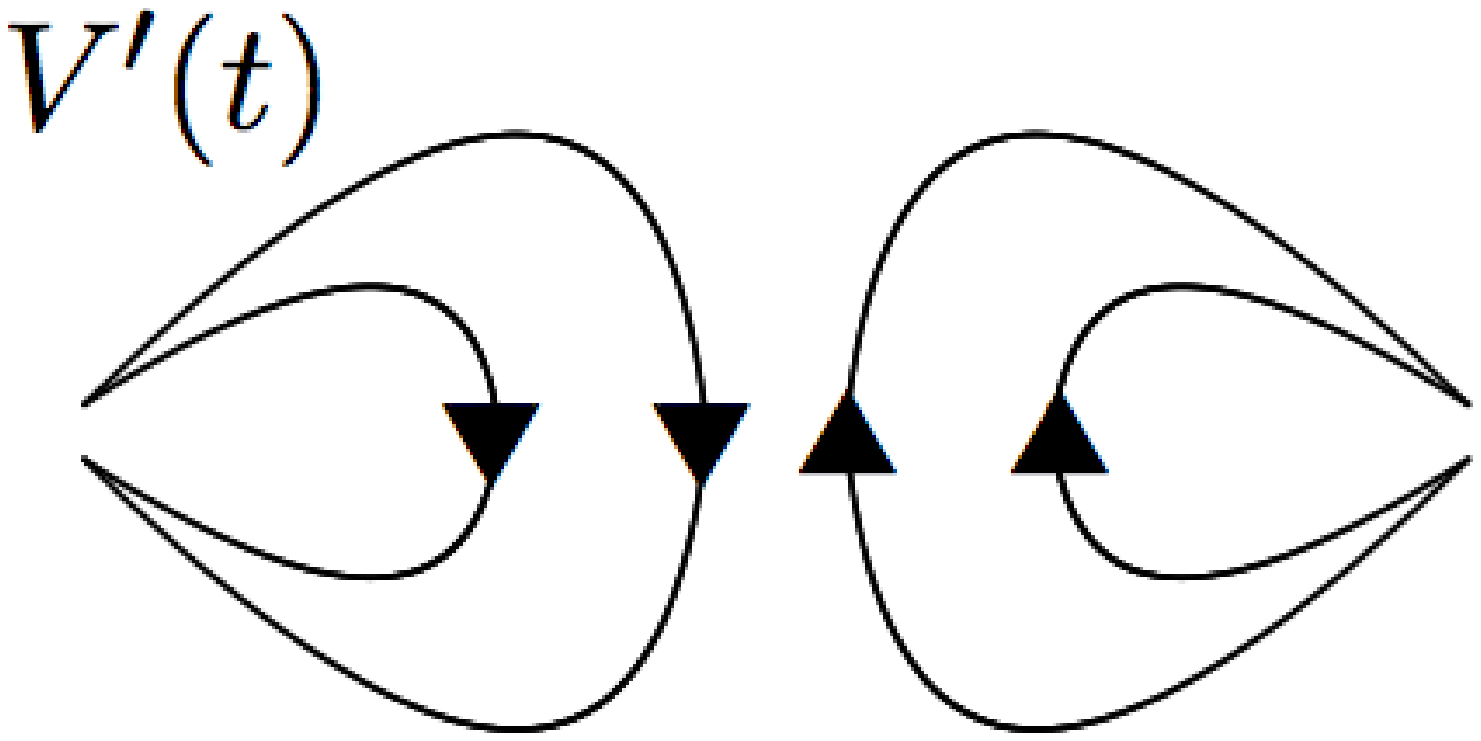}
\caption{
The diagrams for the propagator of the tetraquark operator $G^{\rm{tetra}}(t)$.}
\label{dashDAG}
\end{center}
\end{figure}
where $D^{\prime}(t)$, $A^{\prime}(t)$, and $V^{\prime}(t)$ are shown in Fig.~\ref{dashDAG} 
and their detailed formulas are given in the Appendix. 
The number index of $D^{\prime}(t)$, $A^{\prime}(t)$, and $V^{\prime}(t)$ 
represents the difference of the combination of the color index, 
which is described in detail in the Appendix. 
The difference between Figs.~\ref{DCAG} and \ref{dashDAG} is in the directions of the arrows on the quark lines. 

Both propagators $G^{\rm{molec}}(t)$ and $G^{\rm{tetra}}(t)$ 
contain doubly disconnected diagrams $V(t)$ and $V^{\prime}(t)$, 
which are neglected in our calculations. 
Assuming that the $N_c$ counting scheme\cite{large Nc} also works for $N_c=3$, we apply it to 
the contraction in the diagrams. 
We estimate the orders of the diagrams in Figs.~\ref{DCAG}-\ref{dashDAG}: 
$D(t)$ and  $D^{\prime}(t)$ $\sim {\cal O}(N_c^2)$, 
$C(t)$ $\sim {\cal O}(N_c)$, 
$A(t)$ and $A^{\prime}(t)$ $\sim {\cal O}(N_c)$ and 
$V(t)$ and $V^{\prime}(t)$ $\sim {\cal O}(1)$. 
Under the above assumption, 
we may neglect the doubly disconnected diagrams $V(t)$ and $V^{\prime}(t)$ compared with other diagrams. 
Moreover, the large-$N_c$ counting suggests that the singly disconnected diagrams $A(t)$ and $A^{\prime}(t)$ 
become the same order as the crossed diagram $C(t)$. 
The singly disconnected diagrams may play an essential role in the understanding of four-quark states 
and should not be neglected. 

However, the calculation of the singly disconnected diagrams has a huge computational cost 
because the evaluation of the quark loop on all lattice sites is necessary. 
To reduce the computational time, 
we use the $Z_2$-noise method with the truncated eigenmode approach\cite{McNeile:2000xx,Struckmann:2000bt,Neff:2001zr} 
to estimate the quark loop and evaluate the vacuum expectation value. 
We subtract the contribution of the vacuum expectation value in the singly disconnected 
diagram, which is the same as that from the disconnected diagram of the two-quark operator \cite{Scalar04}. 

\section{Calculated Results}
We generate two-flavor full QCD configurations using the same simulation parameters 
(clover coefficient $C_{SW}=1.68$ and coupling $\beta=1.7$) 
as those in Ref.~\cite{Ali Khan:2000iz}, except for the lattice size. 
The lattice size in our calculation is set to $8^3 \times 16$, which is smaller than that in Ref.~\cite{Ali Khan:2000iz}. 
First we produce the two-flavor full QCD configurations 
using the hybrid Monte Carlo method with the clover-improved Wilson quark action. 
The first 2000 trajectories are updated in the quenched QCD, 
then we switch to simulations with the dynamical fermion. 
The next 100 hybrid Monte Carlo trajectories are discarded for thermalization; 
then we start to store the configurations every ten trajectories. 
The numbers of configurations at the dynamical hopping parameter values of 
$\kappa=0.146$, 0.147, and 0.148 are 16496, 14344, and 11720, respectively. 
Our estimated critical hopping parameter $\kappa_c$ and the lattice size 
are $\kappa_c = 0.152(6)$ and $a=0.269(9)$\, fm, respectively. 
The critical hopping parameter  is estimated by the linear 
extrapolation of the square of the pion mass 
($m_\pi a)^2$ as a function of the inverse of the hopping 
parameter in Fig.\ref{mas_dep}. 
In Fig.\ref{mas_dep} we plot  rho meson masses as a function of the inverse of the 
hopping parameter and compute the value of the rho meson mass 
at the inverse of the critical hopping parameter from the linear extrapolation 
of the plots. 
From comparison between the rho meson mass at $1/\kappa_c$, $m_\rho a$ and 
the physical mass $m_\rho=770$ MeV, we obtain the lattice spacing $a=0.269(9)$ fm.
We list the values of the $\pi$ and $\rho$ meson masses together with the number of configurations at 
$\kappa=0.146, 0.147$, and 0.148 
in Table~\ref{kappa}. 
We calculate the quark propagators using a point source and sink with the clover-improved Wilson quark action. 
For the disconnected diagrams we employ the $Z_2$-noise method with the truncated eigenmode approach. 
We carry out the dilution in the temporal direction\cite{dilution}, 
in which the numbers of noise vectors and eigenvalues are 120 and 12, respectively.

\begin{table}[h]
\caption{
Masses of $\pi$ and $\rho$ and number of configurations.}
\begin{ruledtabular}
\begin{tabular}{cccccc}
$\kappa$ &
$m_{\pi} a$ &
$m_{\pi}$\, MeV &
$m_{\rho} a$ &
$m_{\rho}$\, MeV &
\textrm{Configurations
\footnote{Number of configurations separated from each other by ten trajectories.}}\\
\colrule
0.146 & 1.018(2) & 747(27) & 1.431(4)  & 1050(39) & 16496 \\
0.147 & 0.930(2) & 682(25) & 1.358(6)  &  996(38) & 14344 \\
0.148 & 0.827(4) & 607(23) & 1.304(10) & 956(39) & 11720 \\
\end{tabular}
\end{ruledtabular}
\label{kappa}
\end{table}

\subsection{Importance of the singly disconnected diagrams}

We focus on the importance of the disconnected diagrams in four-quark operators. 
Here we neglect the contribution of the doubly disconnected diagrams in 
the four-quark operators under the assumption that their order is smaller than that of 
other diagrams in the case of large-$N_c$ counting. 
We analyze the propagators of the molecule $G^{\rm molec}$ and 
tetraquark  $G^{\rm tetra}$.

\begin{figure}[h]
\begin{center}
\includegraphics[scale=0.45]{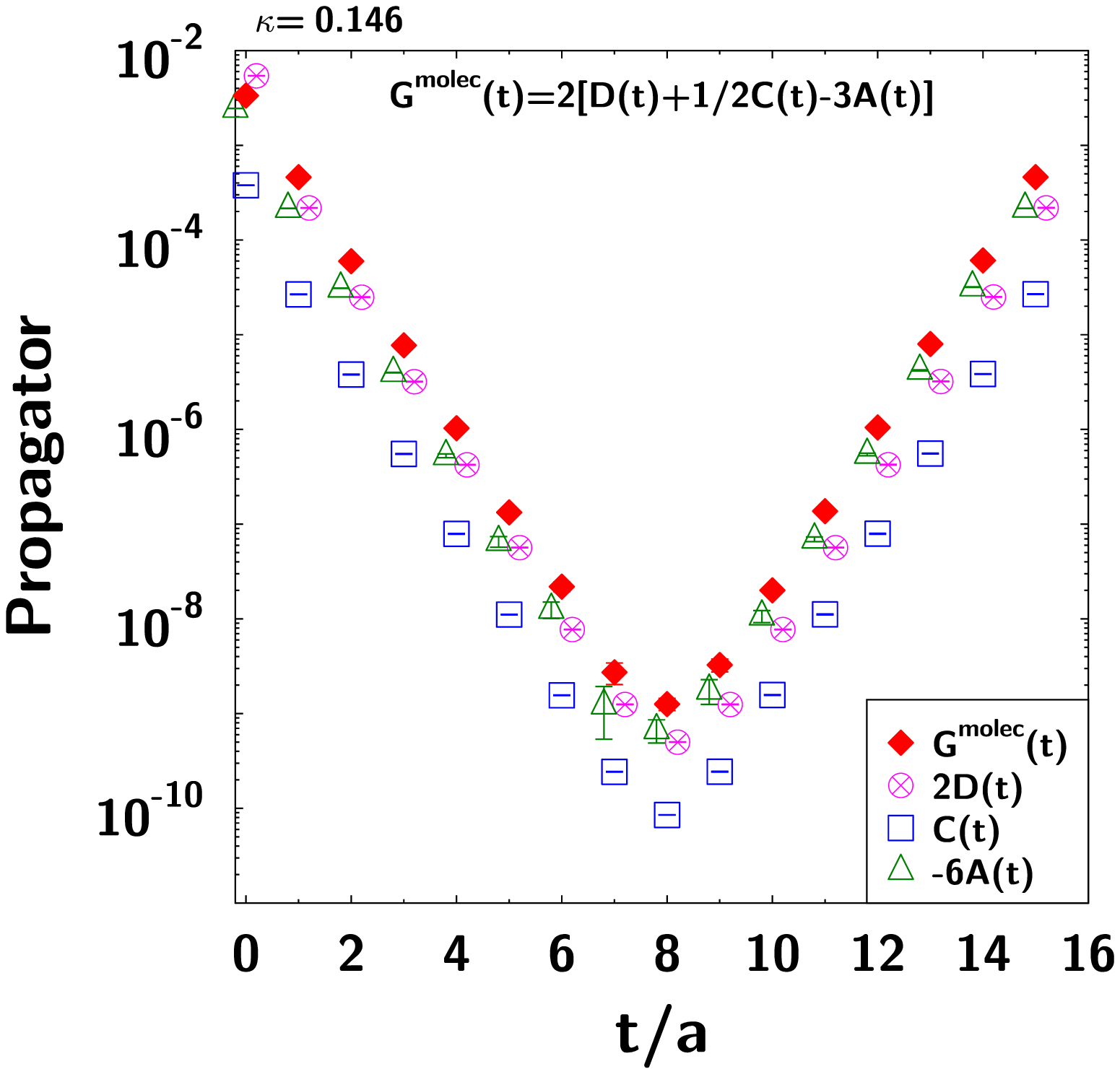}
\includegraphics[scale=0.45]{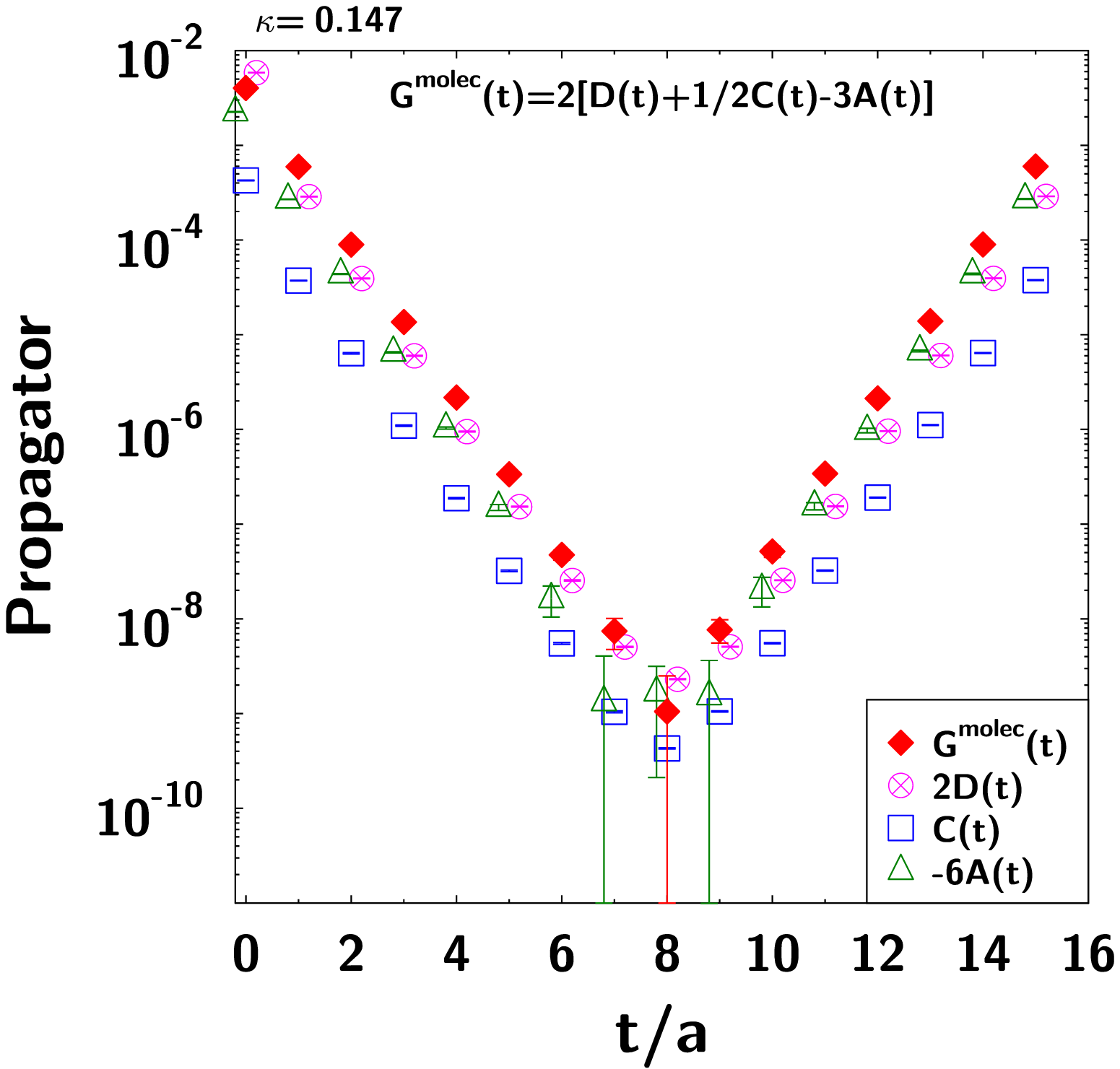}
\includegraphics[scale=0.45]{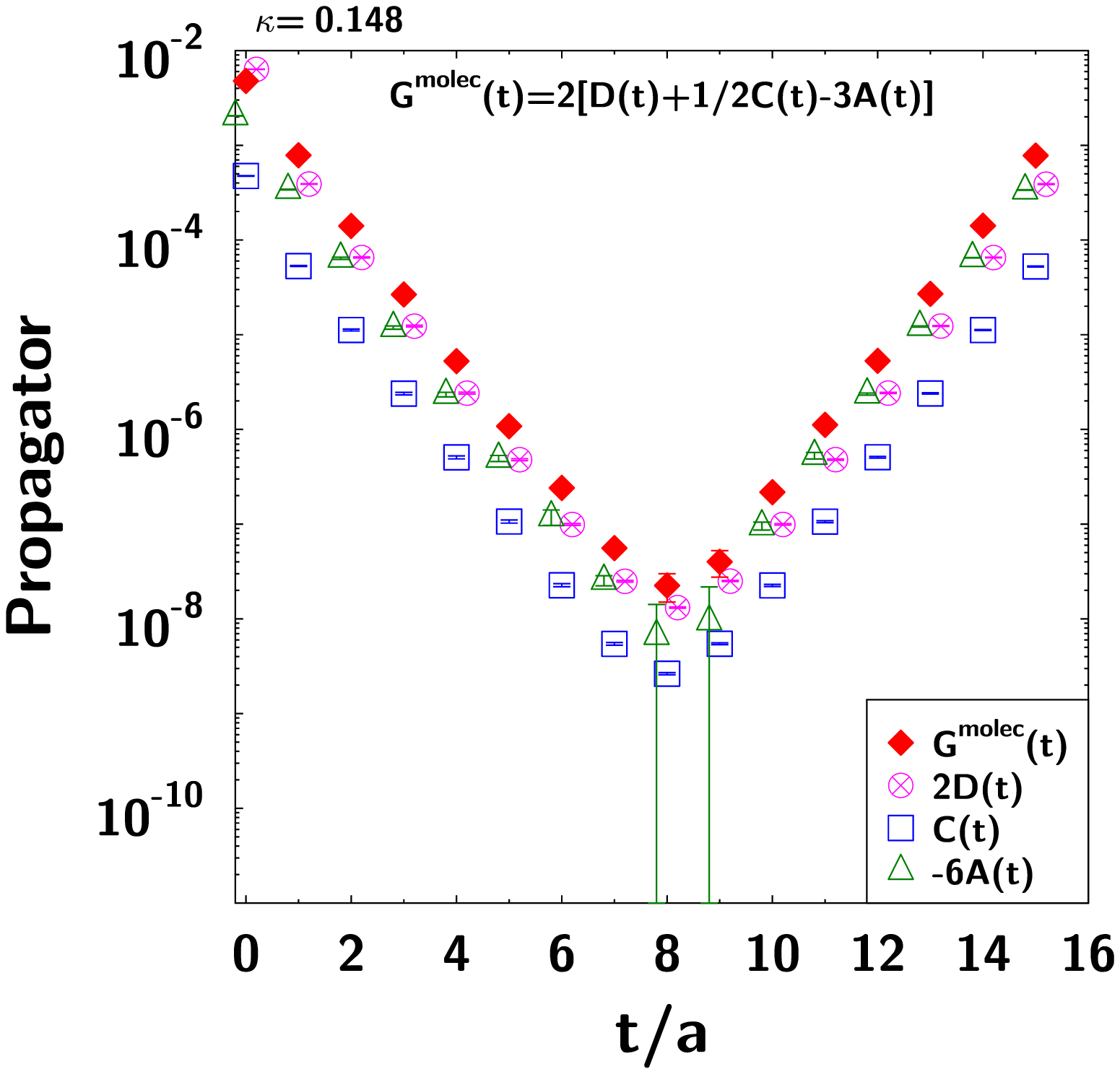}
\caption{ (color online). 
The propagators of the molecule and its components at $\kappa=0.146$, 0.147, and 0.148. 
The solid diamonds represent the propagators of the molecule.  
The open circles, squares, and triangles represent the components of  
the molecular operators $2D(t)$, $C(t)$, and $-6A(t)$ in Eq.~(\ref{molec-molec}), respectively. 
The $D(t)$ and $A(t)$ diagrams are dominant in the molecule. 
The plots of $2D(t)$ and $-6A(t)$ are shifted to $t/a \pm 0.2$ for visibility. 
}
\label{DCA}
\end{center}
\end{figure}

First we show the propagators of the molecular operators at $\kappa=0.146$, 0.147, and 0.148 
in Fig.~\ref{DCA}, together with the propagators of 
diagrams $D(t)$, $C(t)$, and $A(t)$ in Fig.~\ref{DCAG}. 
They are weighted with the coefficients in Eq.~(\ref{molec-molec}) 
to make it clear which diagram is important in the molecule.  
In the propagator of the molecular operator $G^{\rm molec}$ 
the connected diagram $D(t)$ and the singly disconnected diagram $A(t)$ 
are dominant compared with the connected diagram $C(t)$. 
We emphasize that the contribution from the singly disconnected diagram $A(t)$ 
is the same order of magnitude as that from the connected diagram $D(t)$, 
which suggests that the singly disconnected diagram 
should not be neglected in the propagator of the molecule. 

\begin{figure}[h]
\begin{center}
\includegraphics[scale=0.45]{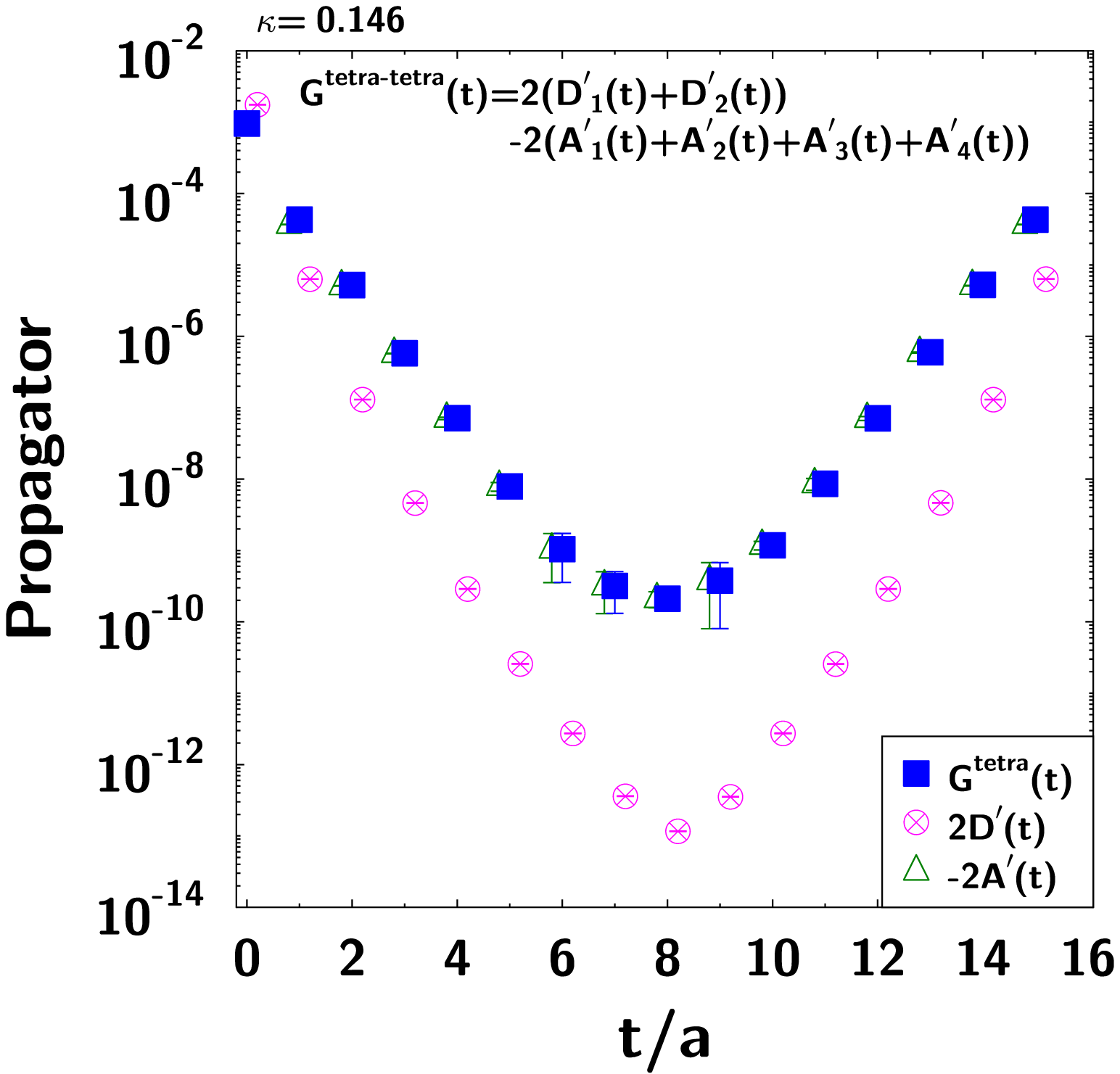}
\includegraphics[scale=0.45]{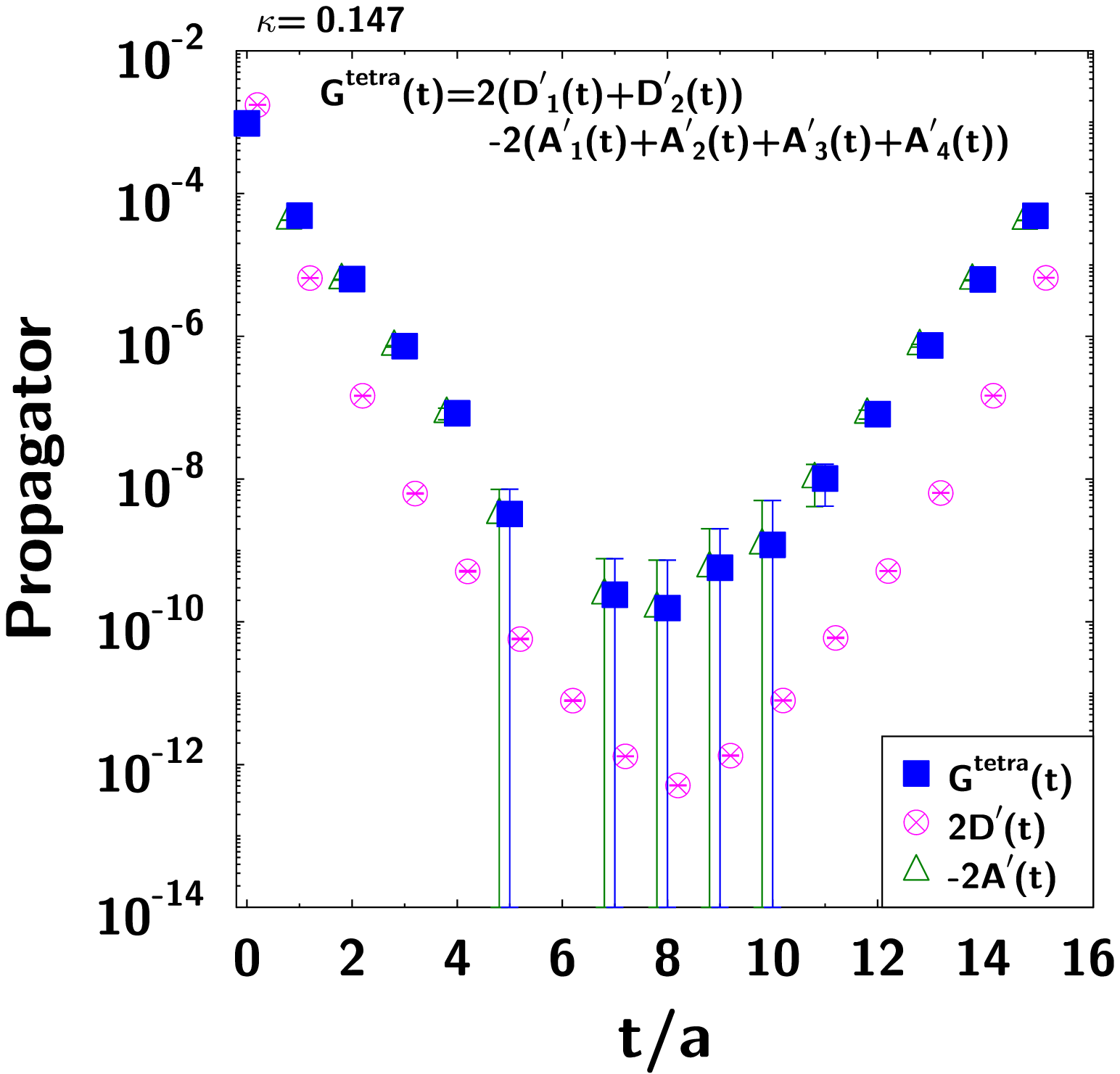}
\includegraphics[scale=0.45]{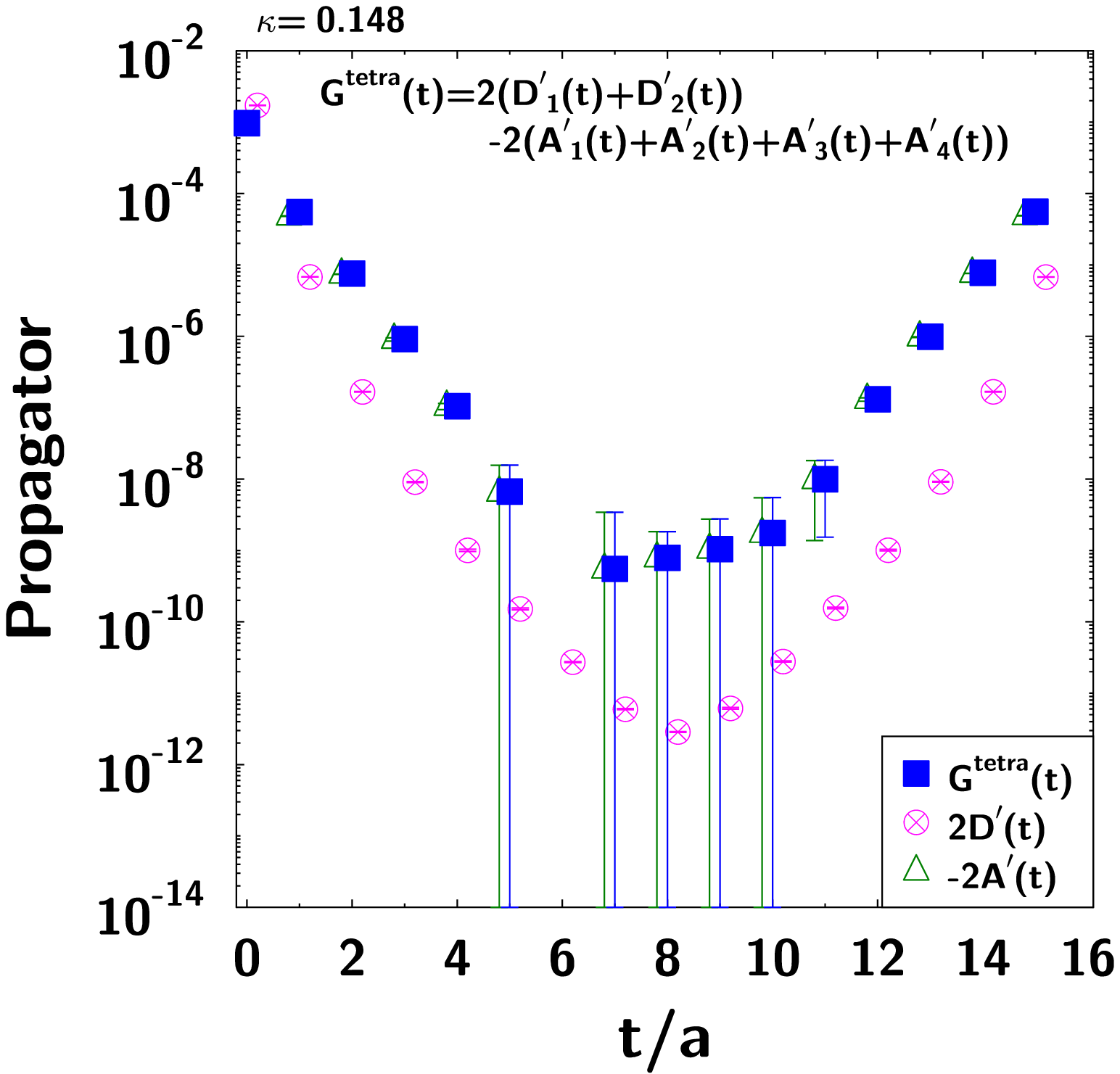}
\caption{ (color online). 
The propagators of the tetra-quark operator and its components at $\kappa=0.146$, 0.147, and 0.148. 
The solid squares represent the propagators of tetraquark. 
The open circles and triangles represent $2D^{\prime}(t)$ and $-2A^{\prime}(t)$, respectively, 
which are plotted at $t/a \pm 0.2$ for visibility. 
The singly disconnected diagrams $A^{\prime}(t)$ are dominant in the tetraquark. 
}
\label{dashDA}
\end{center}
\end{figure}

Next, the propagators of the tetraquark operator at $\kappa=0.146, 0.147$, and 0.148 are shown in Fig.~\ref{dashDA}. 
We also plot the elements of the tetra-quark diagrams 
$D^{\prime}(t)$ and $A^{\prime}(t)$, in Fig.~\ref{dashDAG}, 
where $D^{\prime}(t)$ and $A^{\prime}(t)$ are given by 
$D^{\prime}(t)=D^{\prime}_{1}(t)+D^{\prime}_{2}(t)$ and 
$A^{\prime}(t)=A^{\prime}_{1}(t)+A^{\prime}_{2}(t)+A^{\prime}_{3}(t)+A^{\prime}_{4}(t)$. 
The propagators of the singly disconnected diagrams $A^{\prime}(t)$ at $\kappa=0.147$ and 0.148 have 
some error around $5 \le t \le 10$ in spite of the high-statistics calculation. 
We can see that the main component of the propagator of the tetraquark operator originates 
from the singly disconnected diagrams $A^{\prime}(t)$. 
The absolute values of the propagator of the connected diagrams $D^{\prime}(t)$ are 
much smaller than those of the singly disconnected diagrams $A^{\prime}(t)$. 
We thus cannot neglect the singly disconnected diagram in the investigation of the tetraquark. 
From the comparison between Figs.~\ref{DCA} and \ref{dashDA}, the propagators of the molecule have 
smaller errors than those of the tetraquark. 
The propagators of the molecule have only small errors 
because the interpolation operators of it are composed of two pseudoscalar mesons whose propagators have small errors. 

From Figs.~\ref{DCA}-\ref{dashDA}, 
we see that the singly disconnected diagrams play the key role in understanding the molecule and tetra-quark. 
In particular, 
the singly disconnected diagrams are dominant in the four-quark, 
which is also found in the two-quark state\cite{Scalar04}.

\begin{figure}[h]
\begin{center}
\includegraphics[scale=0.35]{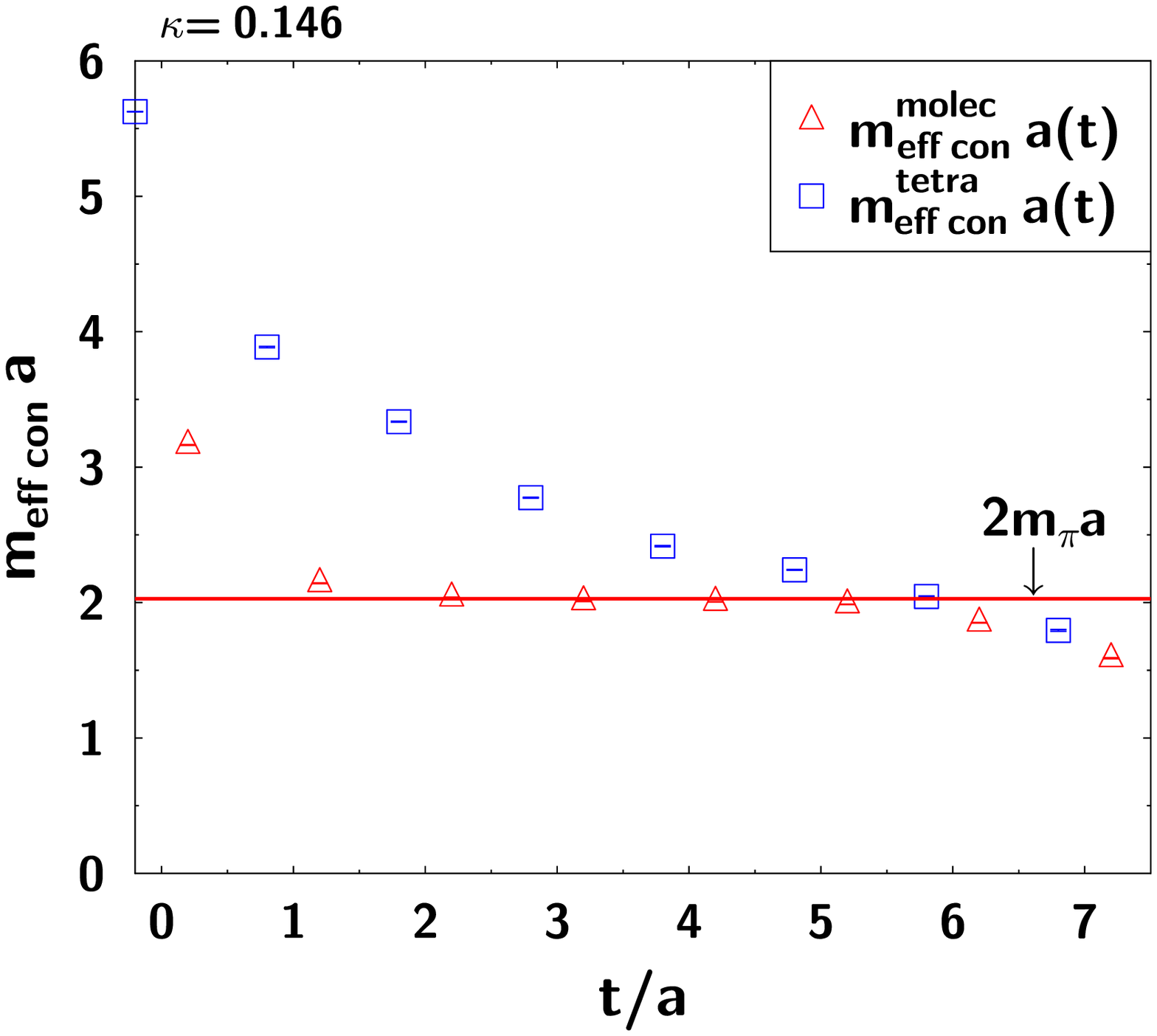}
\includegraphics[scale=0.35]{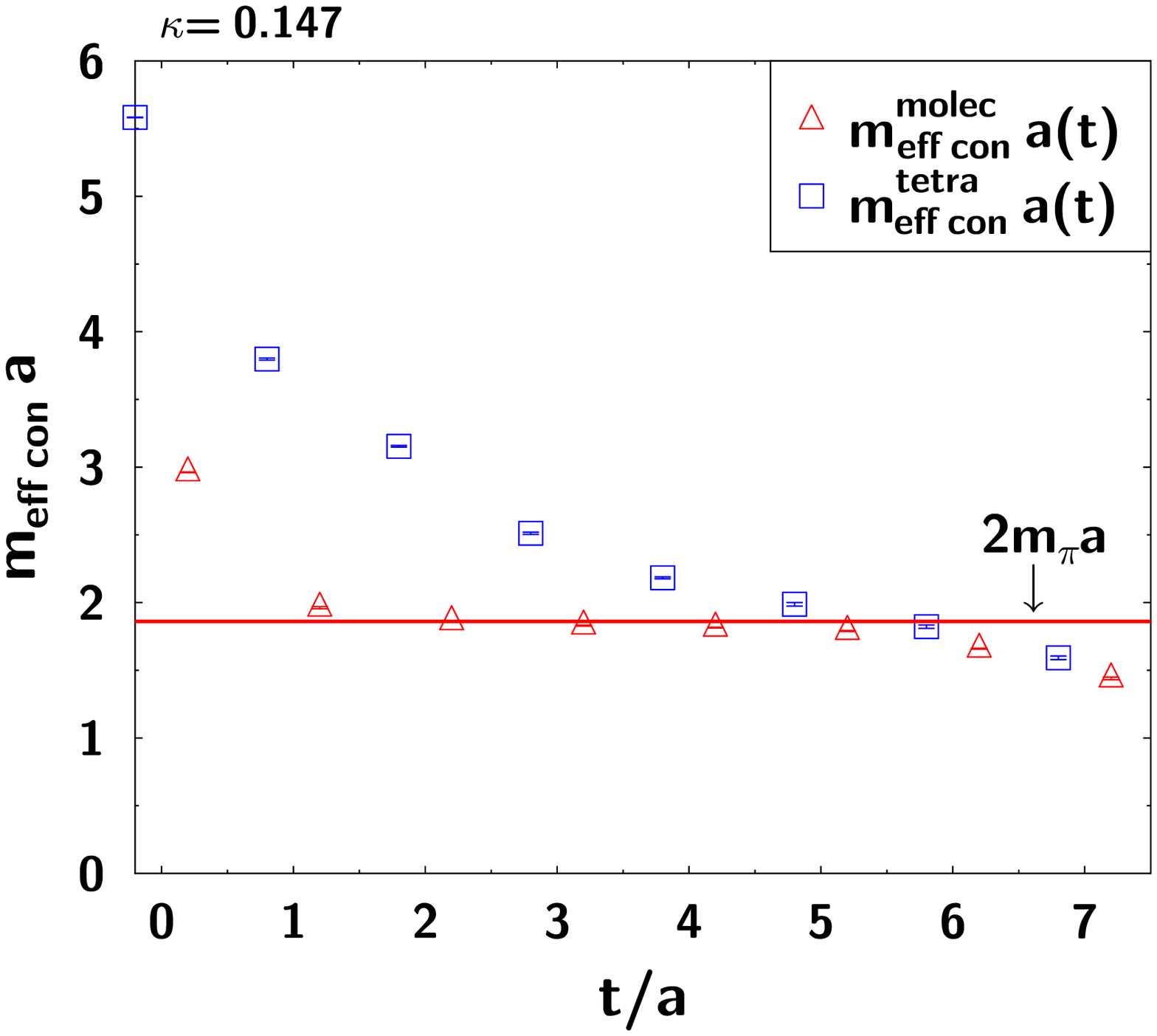}
\includegraphics[scale=0.35]{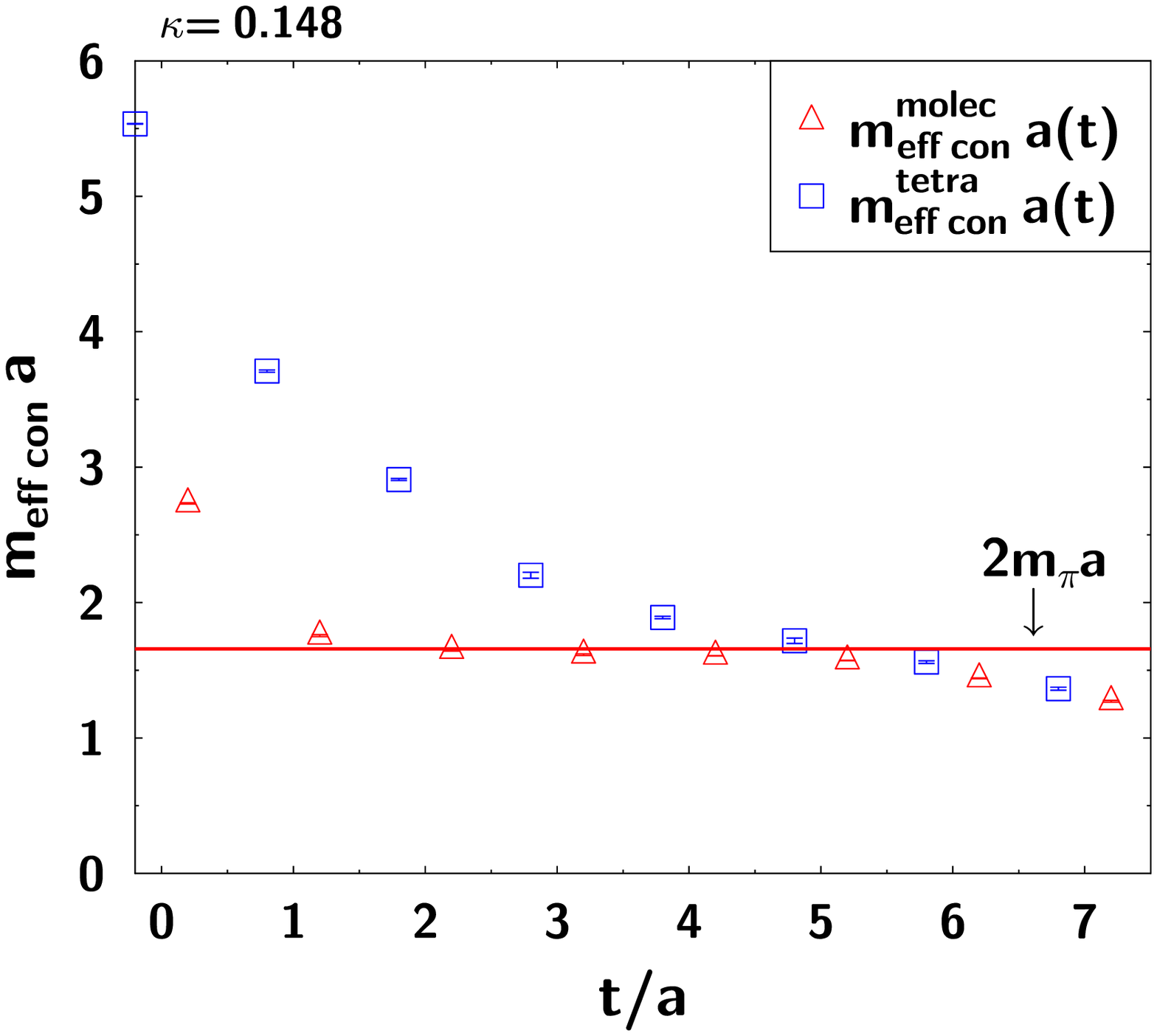}
\caption{(color online). 
The effective masses of the molecule (open triangles) and tetraquark (open squares) 
without the singly disconnected diagrams at $\kappa=0.146$, 0.147, and 0.148. 
The data are plotted at $t/a \pm 0.2$ for visibility. 
}
\label{meffc}
\end{center}
\end{figure}

\begin{figure}[h]
\begin{center}
\includegraphics[scale=0.35]{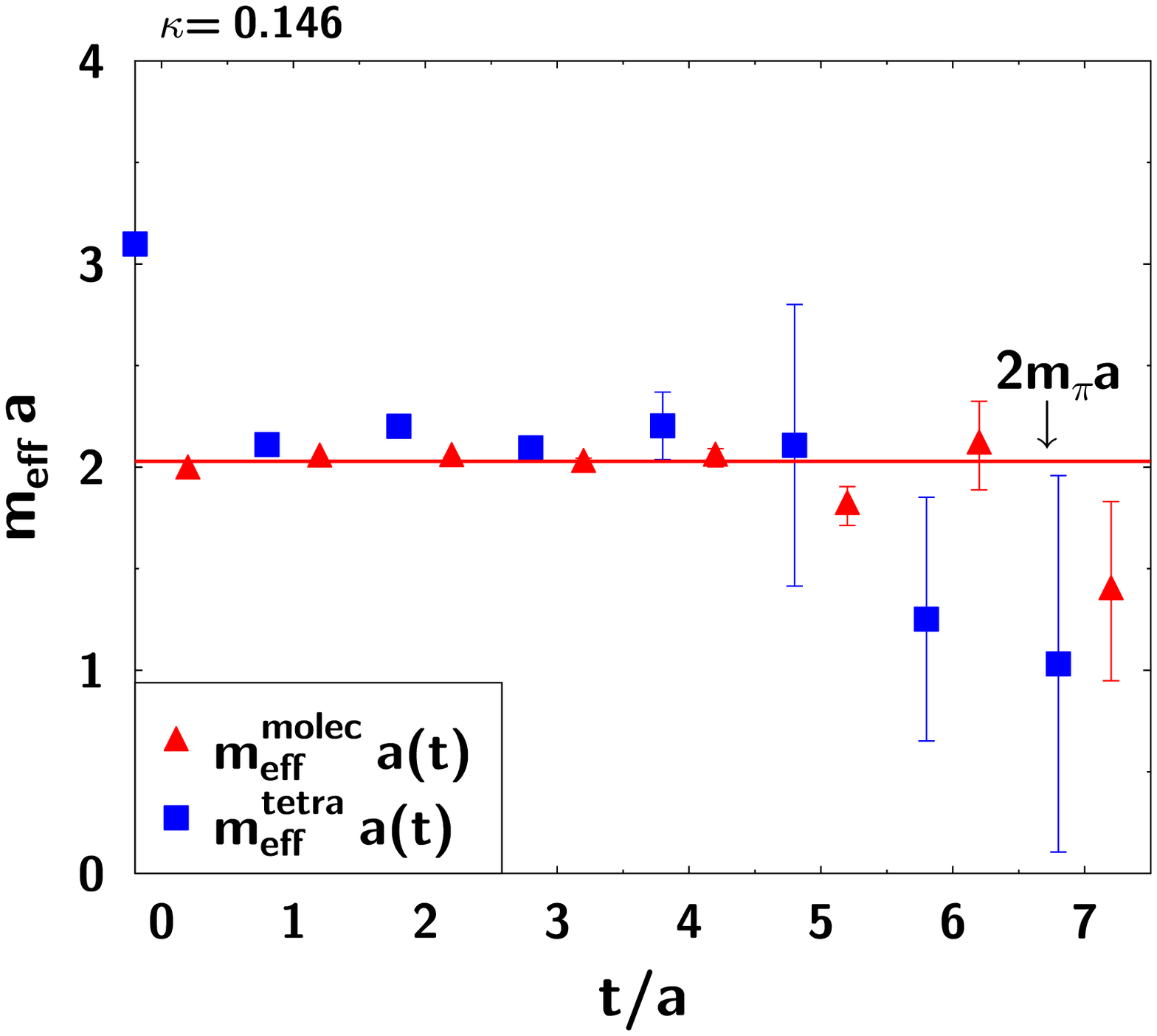}
\includegraphics[scale=0.35]{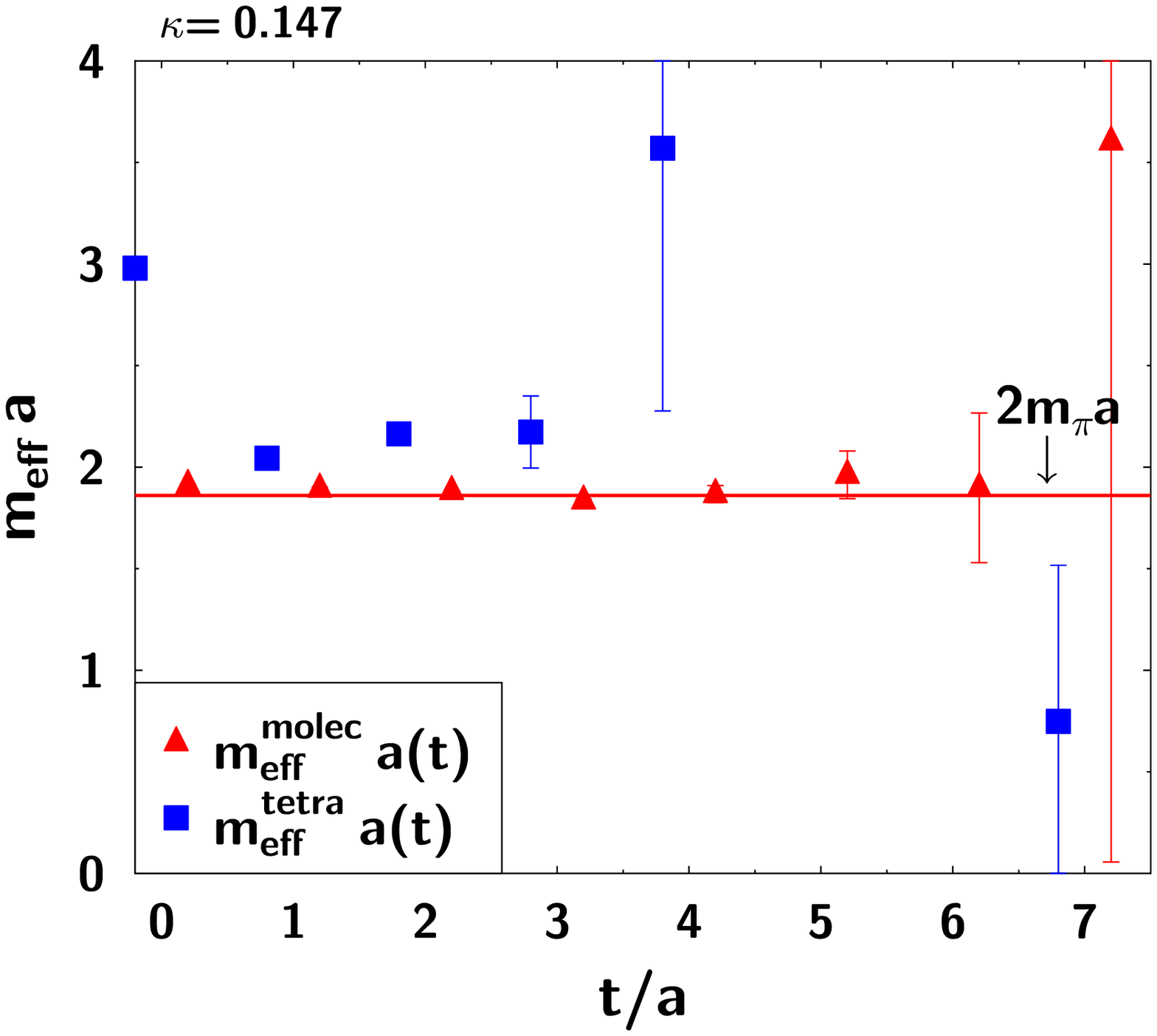}
\includegraphics[scale=0.35]{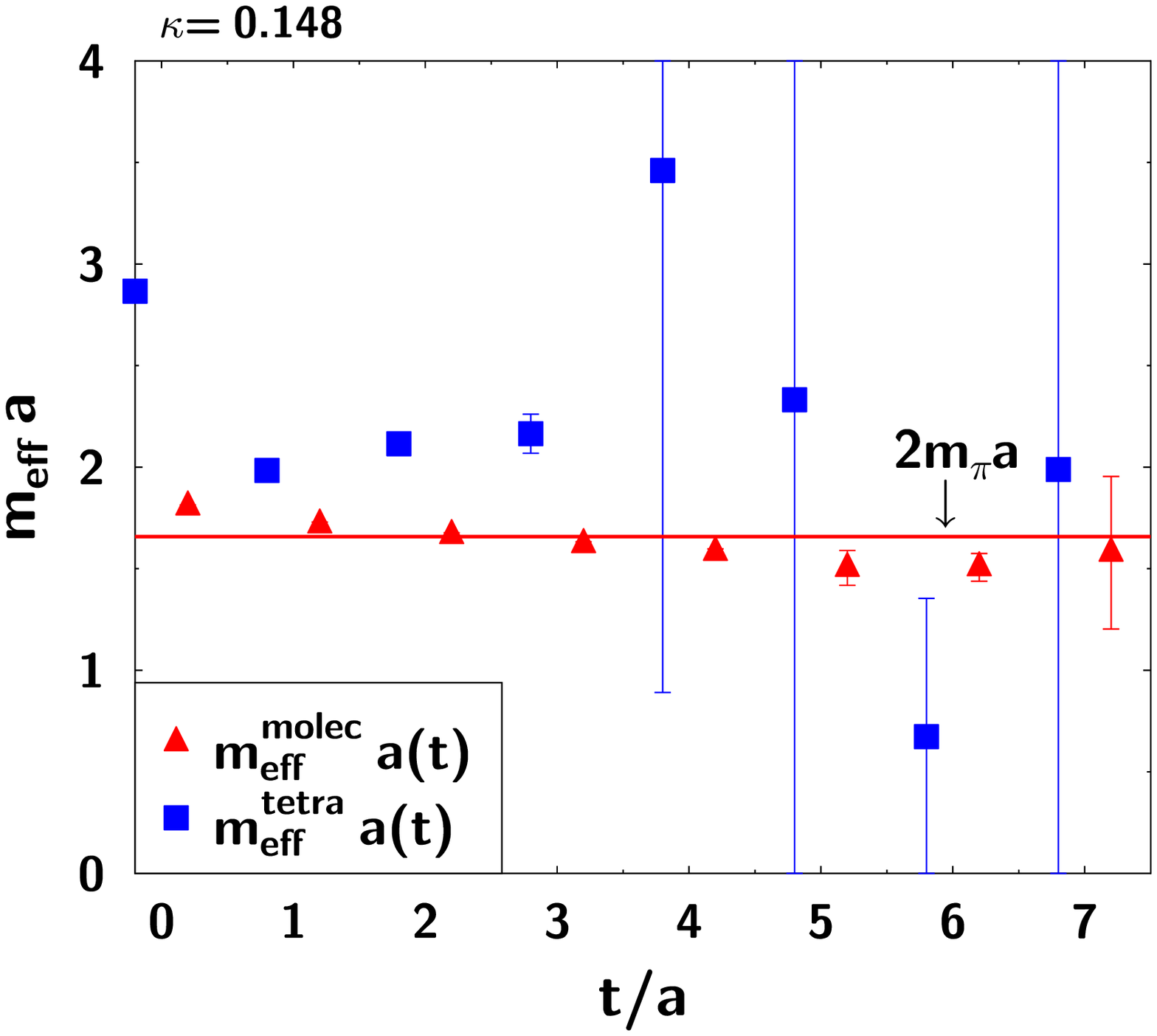}
\caption{(color online). 
The effective masses of the molecule (solid triangles) and tetraquark (solid squares) 
with the singly disconnected diagrams at $\kappa=0.146$, 0.147, and 0.148.
The data are plotted at $t/a \pm 0.2$ for visibility. 
}
\label{meff}
\end{center}
\end{figure}

We show the effective masses 
obtained from the propagators $G^{\rm molec}$
and $G^{\rm tetra}$ in Figs.~\ref{meffc}-\ref{meff}. 
The effective masses are defined by 
\begin{eqnarray}
 \frac{G ^{i} (t)}{G^{i} (t+1)} & = & 
 \frac{e^{-m^i_{\rm eff}(t)t} +  e^{-m^i_{\rm eff}(t)(T-t) }}{e^{-m^i_{\rm eff}(t)(t+1)} +  e^{-m^i_{\rm eff}(t)(T-(t+1)) }}
  , \ 
i = {\rm molec \ or \ tetra}. 
\end{eqnarray}

Figure \ref{meffc} shows the effective masses without the singly disconnected diagrams as a function of time. 
The molecule has a clear plateau in the behavior of the effective masses in the range $1 \le t \le 5$. 
The value of the plateau is the same as $2m_\pi$, which would suggest that the molecule has 
a large overlap with the two-particle $\pi$-$\pi$ scattering state. 
On the other hand, 
the values of effective masses of the tetraquark are larger than 
those of the molecule  at small $t$ and decrease significantly with time as reported in Ref.\cite{Sasa10}. 
We do not observe a clear plateau in the effective masses of the tetraquark. 
At $t=6, 7$ the small mass drop is found  in effective masses of molecule. 
To understand it we need to check whether the small mass drop still exists in the larger lattice size 
calculation. Currently we have not reached any physical interpretation of it. 

\begin{figure}[h]
\begin{center}
\includegraphics[scale=0.5]{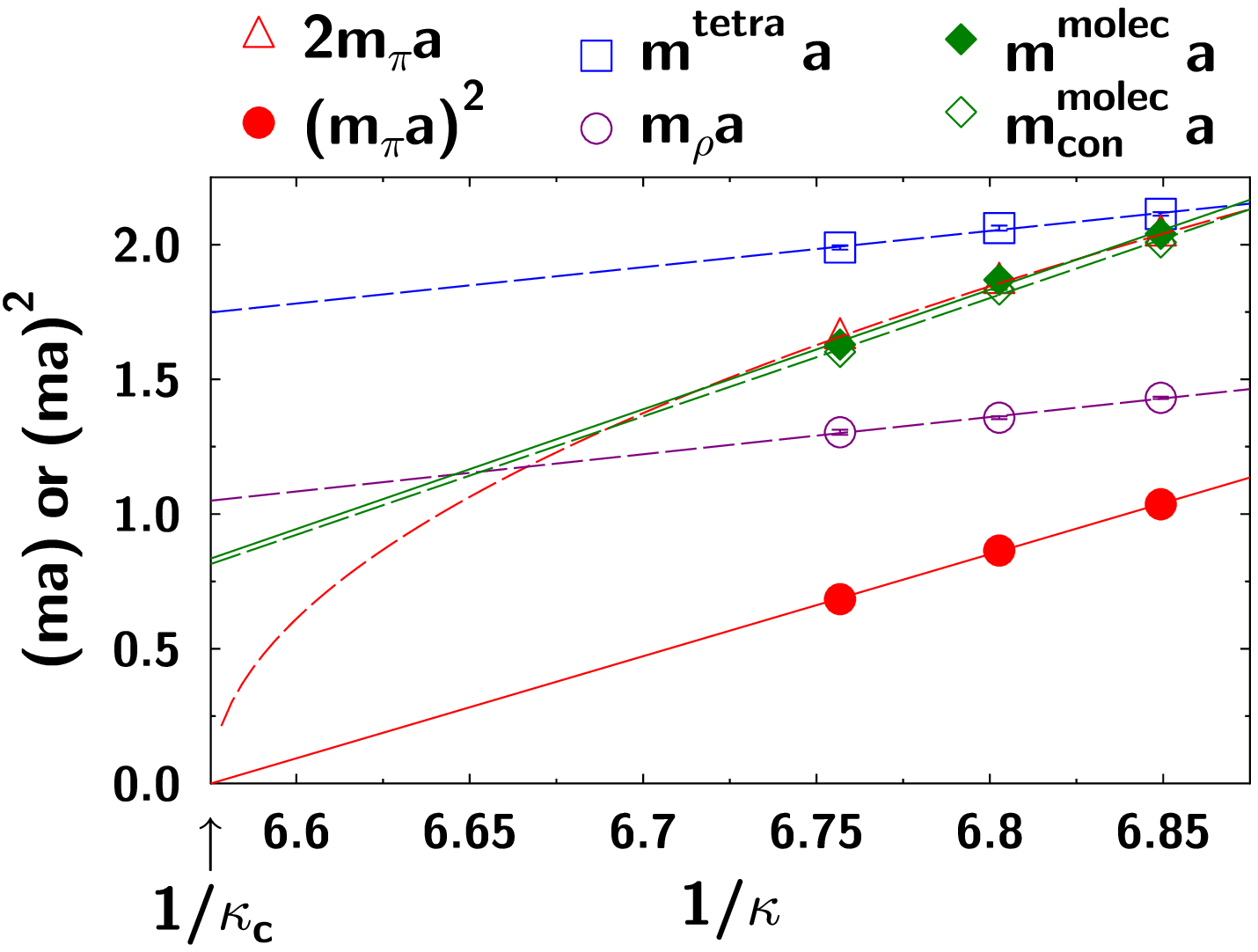}
\caption{(color online). 
The quark mass dependence of 
the square of the $\pi$ meson mass (solid circles), 
double the $\pi$ meson mass (open triangles), 
the $\rho$ meson mass (open circles), 
the mass of the molecule (diamonds), and the mass of the tetra-quark (open squares). 
We plot the masses of the molecule 
both with (solid diamonds) and without (open diamonds) the singly disconnected diagram. 
The chiral limit is given by $\kappa_c=0.152(6)$. 
}
\label{mas_dep}
\end{center}
\end{figure}

In Fig.\ref{meff} we show the effective masses 
as a function of time for the molecule and tetraquark with the singly disconnected diagrams. 
The behavior of the effective masses of the molecule with the singly disconnected diagram 
is almost the same as that without the singly disconnected diagram. 
There is a clear plateau in effective masses whose value is the same as $2m_\pi$. 
We find a dramatic change in the behavior of the effective masses of tetraquark 
due to the existence of the singly disconnected diagrams. 
A plateaulike structure appears at small $t$ whose values are larger than those of 
molecule, which implies that the tetraquark has a small overlap with 
the lowest state in the molecule. 

In Fig.\ref{mas_dep}, 
we display $(m_\pi)^2$, $2m_\pi$, $m_\rho$, $m^{\rm molec}$, 
$m^{\rm molec}_{\rm con}$, and $m^{\rm tetra}$ 
in the lattice unit as a function of the inverse hopping parameter. 
The masses of the molecular operators are obtained from plateaus 
of the effective masses in the range $ 2 \le t \le 5$. 
The difference between the masses of the molecular operator with the singly disconnected 
diagram and those without the singly disconnected diagram is small at $\kappa = 0.146, 0.147$, 
and 0.148. 
In both cases, the extracted masses are identical to $2m_\pi$ which would  suggest that 
the molecular operators have a large overlap with the two-particle $\pi$-$\pi$ scattering state. 
To confirm it, the investigation of the energy shift of the two-particle $\pi$-$\pi$ state 
would be helpful \cite{Jaffe, Luscher}. 
If we assume that the plateaulike structure of the effective masses 
of the tetraquark with the singly disconnected diagrams in $1 \le t \le 4$ exists, 
we can evaluate the mass of the tetraquark.  
The mass from the tetraquark operator is larger than that of the molecular operator and 
the difference between the masses becomes larger at smaller quark mass. 
It  indicates that the tetraquark operators have smaller overlap with the lowest 
state in the molecular operators and the mass of them can be a mixture of the excited state. 
In the calculation, we do not observe any bound four-quark states in the molecular and tetra-quark operators.

\section{Conclusion and outlook}

We investigated the possible significance of the four-quark states in the isosinglet scalar mesons,  
whose quantum numbers are $J^{PC}=0^{++}$, $I=0$, with the two-flavor dynamical quarks on the lattice. 
We reported the results of the propagators and the effective masses of two types of 
interpolation operators for the creation of four-quark states, 
including the estimate of the singly disconnected diagrams, for the first time. 
We showed that the quark loops given by the disconnected diagrams 
play an essential role in propagators of molecular and tetraquark operators. 

We evaluated the effective masses of the molecular operator 
with and without the singly disconnected diagram. 
The difference between the mass of the molecular operator with the singly disconnected diagram 
and that without the singly disconnected diagram is small at $\kappa = 0.146, 0.147$, and 0.148. 
The masses extracted from plateaus in the effective masses of the molecule operator 
with or without the singly disconnected diagram 
are approximately $2 m_{\pi}$, which would suggest that 
the lowest state of  the molecule has large overlap with the two-particle $\pi$-$\pi$ state.

On the other hand, for the tetraquark operator, 
we found that the singly disconnected diagrams markedly affect the effective masses. 
By virtue of the singly disconnected diagrams,  the plateaulike structure appears in the effective masses. 
The value of plateau in effective mass of the tetraquark operators is different from that of molecule operators, which 
implies that the tetraquark operators have smaller overlap with the lowest 
state in the molecule and the mass of them can be a mixture of the excited state. 
The importance of the singly disconnected diagrams in the tetraquark would suggest that the doubly 
disconnected diagrams may be essential for the lowest state in the tetraquark. 
We leave the point for our future work. 

In the current  calculation, we do not observe any  bound four-quark states in molecular and tetraquark operators. 
To reach conclusive results regarding the possible significance of four-quark states in the isosinglet scalar mesons, 
further improvement of the computation is indispensable, such as 
calculation on a larger lattice, 
estimation of the doubly disconnected diagrams in the molecule and tetraquark operators, 
inclusion of other possible interpolation operators for the four-quark states, 
and the use of the variational method with possible interpolation operators. 
In particular, to investigate the existence of a pole, 
the calculation of all diagrams in the four-quark operators is indispensable. 
The evaluation of the doubly disconnected diagrams is important 
because they make the scattering amplitudes unitary\cite{Sharpe:1992pp}. 

In addition to the four-quark states, there are many possible states 
with the same quantum number as that of the isosinglet scalar mesons: 
two-quark states $(\bar q q)$, glueballs $(gg)$, hybrid states $(\bar q q g)$, and so on.  
For the comprehensive understanding of the isosinglet scalar mesons, 
these interpolation operators should be taken into account. 
We carried out the calculation with heavy quark masses of $m_{\pi}=607, 682,$ and 747\, MeV, which 
are far from the physical $\pi$ mass. 
Computation with light quark masses close to the physical point 
would change the features of the molecule and tetraquark. 

\begin{acknowledgments}
This work was supported in part by 
the Nagoya University Program for Leading Graduate Schools 
"Leadership Development Program for Space Exploration and Research", 
Grant-in-Aid for Scientific Research (S) (Grant No. 22224003), 
the Kurata Memorial Hitachi Science and Technology Foundation, 
and the Daiko Foundation. 
This work was partially supported by Grants-in-Aid for 
Research Activity of Matsumoto University (Grant No. 14111048). 
This work was supported by Grants-in-Aid 
for Scientific Research (Kakenhi) Grants No. 24340054, No. 26610072 and No. 15H03663. 
The simulation was performed on an NEC SX-9 and SX-ACE supercomputers at RCNP, Osaka University, 
and was conducted using the Fujitsu PRIMEHPC FX10 System (Oakleaf-FX, Oakbridge-FX) 
in the Information Technology Center, The University of Tokyo. 
\end{acknowledgments}

\appendix

\section{Propagators of the four-quark states}
The detailed descriptions of diagrams $D(t)$, $C(t)$, $A(t)$, and $V(t)$ in Eq.~(\ref{molec-molec}) are given by 
\begin{widetext}
\begin{eqnarray}
    D(t) &=& \Big\langle 
                  \sum_{\mbf{x},\mbf{y} \ a,b,c,d} \!\!\! 
                     \Tr \kakk{\gamma_5 \qprop^{ac}_{t,\mbf{x};0,\mbf{z}}
                                     \gamma_5 \qprop^{ca}_{0,\mbf{z};t,\mbf{x}}}
                      \Tr \kakk{\gamma_5 \qprop^{bd}_{t,\mbf{y};0,\mbf{w}}
                                     \gamma_5 \qprop^{db}_{0,\mbf{w};t,\mbf{y}}} \Big\rangle, \label{Eq-a1}\\ 
  C(t) &=&   \Big\langle 
	                  \sum_{\mbf{x},\mbf{y} \ a,b,c,d} \!\!\! 
                     \Tr \kakk{\gamma_5 \qprop^{ac}_{t,\mbf{x};0,\mbf{z}}
                                    \gamma_5 \qprop^{cb}_{0,\mbf{z};t,\mbf{y}} 
                                    \gamma_5 \qprop^{bd}_{t,\mbf{y};0,\mbf{w}}
                                    \gamma_5 \qprop^{da}_{0,\mbf{w};t,\mbf{x}}} \Big\rangle, \\ 
  A(t) &=&   \Big\langle 
	        \sum_{\mbf{x},\mbf{y} \ a,b,c,d} \!\!\! 
                 \Tr \kakk{\gamma_5 \qprop^{ab}_{t,\mbf{x};t,\mbf{y}}
                                \gamma_5 \qprop^{bc}_{t,\mbf{y};0,\mbf{z}} 
                                \gamma_5 \qprop^{cd}_{0,\mbf{z};0,\mbf{w}}
                                \gamma_5 \qprop^{da}_{0,\mbf{w};t,\mbf{x}}} \Big\rangle, \\ 
  V(t) &=&   \Big\langle 
	        \sum_{\mbf{x},\mbf{y} \ a,b,c,d} \!\!\! 
                 \Tr \kakk{\gamma_5 \qprop^{ab}_{t,\mbf{x};t,\mbf{y}}
                                \gamma_5 \qprop^{ba}_{t,\mbf{y};t,\mbf{x}}}
                  \Tr \kakk{\gamma_5 \qprop^{cd}_{0,\mbf{z};0,\mbf{w}}
                                 \gamma_5 \qprop^{dc}_{0,\mbf{w};0,\mbf{z}}}  \Big\rangle ,\label{Eq-a4}\ \ \ \ \ \ 
\end{eqnarray}
\end{widetext}
where the brackets represent the functional integral over gauge configurations and $\Tr$ operates 
on the Dirac spinor. 
Because the mass difference between the up and down quarks is neglected the propagators of 
each of them are expressed by $W^{-1}$. 
In Eqs.~(\ref{Eq-a1})-(\ref{Eq-a4}), the superscripts $a$, $b$, $c$, and $d$ of $W^{-1}$ 
are the indeices of the color. 
We put the source points at $w$ and $z$ in the calculation of the propagators and 
sum over sink points $x$ and $y$. 

The explicit descriptions of $D^{\prime}(t)$, $A^{\prime}(t)$, and $V^{\prime}(t)$ in Eq.~(\ref{tetra-tetra}) 
are written by 
\begin{eqnarray}
 D^{\prime}_1(t) &=& \Big\langle \sum_{\mbf{x},\mbf{y} \ b,c,d,e} \!\!\!\!\!
                                  \Tr \kakk{\ka{C\gamma_5}^T \qprop  ^{bd}_{t,\mbf{x};0,\mbf{z}}
                                                       C\gamma_5 \ka{W^{-1T}}^{ec}_{0,\mbf{z};t,\mbf{x}} } \non \\
                          & & \ \ \ \ \ \ \ \ \ \ \ 
                                  \Tr \kakk{\ka{C\gamma_5}^T \qprop  ^{db}_{0,\mbf{w};t,\mbf{y}} 
                                                       C\gamma_5 \ka{W^{-1T}}^{ce}_{t,\mbf{y};0,\mbf{w}}} \Big\rangle,  \\
 D^{\prime}_2(t) &=& \Big\langle \sum_{\mbf{x},\mbf{y} \ b,c,d,e} \!\!\!\!\!
                                  \Tr \kakk{\ka{C\gamma_5}^T \qprop  ^{bd}_{t,\mbf{x};0,\mbf{z}} 
                                                       C\gamma_5 \ka{W^{-1T}}^{ec}_{0,\mbf{z};t,\mbf{x}} } \non \\
                          & & \ \ \ \ \ \ \ \ \ \ \ 
                                  \Tr \kakk{\ka{C\gamma_5}^T \qprop  ^{dc}_{0,\mbf{w};t,\mbf{y}} 
                                                 \ka{C\gamma_5}^T \ka{W^{-1T}}^{be}_{t,\mbf{y};0,\mbf{w}} } \Big\rangle,  \\
 A^{\prime}_1(t) &=& \Big\langle \sum_{\mbf{x},\mbf{y} \ b,c,d,e} \!\!\!\!\!
                                  \Tr \Big[ \ka{C\gamma_5}^T \qprop  ^{bb}_{t,\mbf{x};t,\mbf{y}} 
                                                       C\gamma_5 \ka{W^{-1T}}^{cd}_{t,\mbf{y};0,\mbf{w}}  \non \\
                          & & \ \ \ \ \ \ \ \ \ \ \ \ \ \ \ \ 
                                                       C\gamma_5 \qprop  ^{ee}_{0,\mbf{w};0,\mbf{z}}
                                                  \ka{C\gamma_5}^T \ka{W^{-1T}}^{dc}_{0,\mbf{z};t,\mbf{x}} \Big] \Big\rangle, \\
 A^{\prime}_2(t) &=& \Big\langle \sum_{\mbf{x},\mbf{y} \ b,c,d,e} \!\!\!\!\!
                                  \Tr \Big[ \ka{C\gamma_5}^T \qprop  ^{bb}_{t,\mbf{x};t,\mbf{y}} 
                                                  C\gamma_5 \ka{W^{-1T}}^{ce}_{t,\mbf{y};0,\mbf{w}}    \non \\
                          & & \ \ \ \ \ \ \ \ \ \ \ \ \ \ \ \ 
                                                  \ka{C\gamma_5}^T \qprop  ^{de}_{0,\mbf{w};0,\mbf{z}} 
                                                  \ka{C\gamma_5}^T \ka{W^{-1T}}^{dc}_{0,\mbf{z};t,\mbf{x}} \Big] \Big\rangle, \\
 A^{\prime}_3(t) &=& \Big\langle \sum_{\mbf{x},\mbf{y} \ b,c,d,e} \!\!\!\!\!
                                  \Tr \Big[ \ka{C\gamma_5}^T \qprop  ^{bc}_{t,\mbf{x};t,\mbf{y}}
                                                  \ka{C\gamma_5}^T \ka{W^{-1T}}^{bd}_{t,\mbf{y};0,\mbf{w}}    \non \\
                          & & \ \ \ \ \ \ \ \ \ \ \ \ \ \ \ \ 
                                                  C\gamma_5 \qprop  ^{ee}_{0,\mbf{w};0,\mbf{z}} 
                                                  \ka{C\gamma_5}^T \ka{W^{-1T}}^{dc}_{0,\mbf{z};t,\mbf{x}} \Big] \Big\rangle, \\
 A^{\prime}_4(t) &=& \Big\langle \sum_{\mbf{x},\mbf{y} \ b,c,d,e} \!\!\!\!\!
                                  \Tr \Big[ \ka{C\gamma_5}^T \qprop  ^{bc}_{t,\mbf{x};t,\mbf{y}}
                                                  \ka{C\gamma_5}^T \ka{W^{-1T}}^{be}_{t,\mbf{y};0,\mbf{w}}    \non \\
                          & & \ \ \ \ \ \ \ \ \ \ \ \ \ \ \ \ 
                                                  \ka{C\gamma_5}^T \qprop  ^{de}_{0,\mbf{w};0,\mbf{z}}  
                                                  \ka{C\gamma_5}^T \ka{W^{-1T}}^{dc}_{0,\mbf{z};t,\mbf{x}} \Big] \Big\rangle, \\
 V^{\prime}_1(t) &=& \Big\langle  \sum_{\mbf{x},\mbf{y} \ b,c,d,e} \!\!\!\!\!
                                  \Tr \kakk{\ka{C\gamma_5}^T \qprop  ^{bb}_{t,\mbf{x};t,\mbf{y}}
                                                      C\gamma_5 \ka{W^{-1T}}^{cc}_{t,\mbf{y};t,\mbf{x}} }  \non \\
                          & & \ \ \ \ \ \ \ \ \ \ \ 
                                 \Tr \kakk{\ka{C\gamma_5}^T \qprop  ^{dd}_{0,\mbf{w};0,\mbf{z}}  
                                                     C\gamma_5 \ka{W^{-1T}}^{ee}_{0,\mbf{z};0,\mbf{w}}}\Big\rangle, \\
 V^{\prime}_2(t) &=& \Big\langle  \sum_{\mbf{x},\mbf{y} \ b,c,d,e} \!\!\!\!\!
                                   \Tr \kakk{\ka{C\gamma_5}^T \qprop ^{bb}_{t,\mbf{x};t,\mbf{y}} 
                                                        C\gamma_5 \ka{W^{-1T}}^{cc}_{t,\mbf{y};t,\mbf{x}} (y,x) }  \non \\
                          & & \ \ \ \ \ \ \ \ \ \ \ 
                                   \Tr \kakk{\ka{C\gamma_5}^T \qprop  ^{de}_{0,\mbf{w};0,\mbf{z}}
                                                  \ka{C\gamma_5}^T \ka{W^{-1T}}^{de}_{0,\mbf{z};0,\mbf{w}} } \Big\rangle, \\
 V^{\prime}_3(t) &=& \Big\langle \sum_{\mbf{x},\mbf{y} \ b,c,d,e} \!\!\!\!\!
                                  \Tr \kakk{\ka{C\gamma_5}^T \qprop  ^{bc}_{t,\mbf{x};t,\mbf{y}}  
                                                  \ka{C\gamma_5}^T \ka{W^{-1T}}^{bc}_{t,\mbf{y};t,\mbf{x}} }  \non \\
                          & & \ \ \ \ \ \ \ \ \ \ \ 
                                  \Tr \kakk{\ka{C\gamma_5}^T \qprop  ^{dd}_{0,\mbf{w}0,\mbf{z}} 
                                                  C\gamma_5 \ka{W^{-1T}}^{ee} _{0,\mbf{z};0,\mbf{w}} } \Big\rangle, \\
 V^{\prime}_4(t)&=& \Big\langle \sum_{\mbf{x},\mbf{y} \ b,c,d,e} \!\!\!\!\!
                                 \Tr \kakk{\ka{C\gamma_5}^T \qprop  ^{bc}_{t,\mbf{x};t,\mbf{y}} 
                                                \ka{C\gamma_5}^T \ka{W^{-1T}}^{bc}_{t,\mbf{y};t,\mbf{x}}}  \non \\
                          & & \ \ \ \ \ \ \ \ \ \ \ 
                                 \Tr \kakk{\ka{C\gamma_5}^T \qprop  ^{de}_{0,\mbf{w};0,\mbf{z}}
                                                \ka{C\gamma_5}^T \ka{W^{-1T}}^{de}_{0,\mbf{z};0,\mbf{w}} } \Big \rangle, \ \ \ 
\end{eqnarray}
where $C$ is the charge conjugation matrix.

\bibliography{apssamp}

\begin{thebibliography}{99}

\bibitem{Weinberg2} 
S.~Weinberg, {\it The Quantum Theory of Fields}, Vol.\ II, (Cambridge University Press, Cambridge, 1999).

\bibitem{PDG14} 
  K.~A.~Olive {\it et al.} [Particle Data Group Collaboration], 
  Chin.\ Phys.\ C {\bf 38}, 090001 (2014).

\bibitem{Jaffe1} R.~L.~Jaffe, Phys. Rev. D {\bf 15}, 267 (1977);
                        \ R.~L.~Jaffe, Phys. Rev. D {\bf 15}, 281 (1977).

\bibitem{X3872} 
  S.-K.~Choi {\it et al.}  [Belle Collaboration],
  Phys.\ Rev.\ Lett. {\bf 91}, 262001 (2003).

\bibitem{Y4260} 
  C.Z.~Yuan {\it et al.}  [Belle Collaboration],
  Phys.\ Rev.\ Lett. {\bf 99}, 182004 (2007).

\bibitem{Z4430} 
  S.-K.~Choi {\it et al.}  [Belle Collaboration],
  Phys.\ Rev.\ Lett. {\bf 100}, 142001 (2008).

\bibitem{Zb10610_Zb10650} 
  A.~Bondar {\it et al.}  [Belle Collaboration],
  Phys.\ Rev.\ Lett. {\bf 108}, 122001 (2012).

\bibitem{Jaffe} 
  M.~G.~Alford and R.~L.~Jaffe, 
  Nucl.\ Phys.\ B {\bf 578}, 367-382 (2000)
  [hep-lat/0001023].

\bibitem{Suganuma:2007uv} 
  H.~Suganuma, K.~Tsumura, N.~Ishii and F.~Okiharu,
  Prog.\ Theor.\ Phys.\ Suppl.\  {\bf 168}, 168 (2007)
  [arXiv:0707.3309 [hep-lat]].

\bibitem{Mathur:2006bs} 
  N.~Mathur, A.~Alexandru, Y.~Chen, S.~J.~Dong, T.~Draper, I.~Horvath, F.~X.~Lee, K.~F.~Liu, S.~Tamhankar, and J.~B.~Zhang,
  Phys.\ Rev.\ D {\bf 76}, 114505 (2007)
  [hep-ph/0607110].

\bibitem{Loan:2008sd} 
  M.~Loan, Z.~H.~Luo and Y.~Y.~Lam,
  Eur.\ Phys.\ J.\ C {\bf 57}, 579 (2008)
  [arXiv:0907.3609 [hep-lat]].

\bibitem{Sasa09} 
  S.~Prelovsek and D.~Mohler,
  Phys.\ Rev.\ D {\bf 79}, 014503 (2009).

\bibitem{Scalar04} 
  T.~Kunihiro, S.~Muroya, A.~Nakamura, C.~Nonaka, M.~Sekiguchi, and H.~Wada [SCALAR Collaboration],
  Phys.\ Rev.\ D {\bf 70}, 034504 (2004)
  [hep-ph/0310312].

\bibitem{McNeile:2000xx} 
  C.~McNeile and C.~Michael  [UKQCD Collaboration],
  Phys.\ Rev.\ D {\bf 63}, 114503 (2001)
  [hep-lat/0010019].

\bibitem{Struckmann:2000bt} 
  T.~Struckmann {\it et al.}  [TXL and T(X)L Collaborations],
  Phys.\ Rev.\ D {\bf 63}, 074503 (2001)
  [hep-lat/0010005].

\bibitem{Neff:2001zr} 
  H.~Neff, N.~Eicker, T.~Lippert, J.~W.~Negele and K.~Schilling,
  Phys.\ Rev.\ D {\bf 64}, 114509 (2001)
  [hep-lat/0106016].

\bibitem{UKQCD06_1} 
  A.~Hart, C.~McNeile, C.~Michael, and J.~Pickavance [UKQCD Collaboration],
  Phys.\ Rev.\ D {\bf 74}, 114504 (2006).

\bibitem{Scalar07} 
  H.~Wada, T.~Kunihiro, S.~Muroya, A.~Nakamura, C.~Nonaka, and M.~Sekiguchi [SCALAR Collaboration],
  Phys.\ Let.\ B {\bf 652}, 250 (2007).

\bibitem{UKQCD06_2} 
  C.~McNeile and C.~Michael [UKQCD Collaboration],
  Phys.\ Rev.\ D {\bf 74}, 014508 (2006).

\bibitem{BGR12} 
  G.~P.~Engel, C.~B.~Lang, M.~Limmer, D.~Mohler, and A.~Sch$\ddot{\rm a}$fer [BGR Collaboration],
  Phys.\ Rev.\ D {\bf 85}, 034508 (2012).

\bibitem{ETM13} 
  C.~Alexandrou, J.~O.~Daldrop, M.~D.~Brida, M.~Gravina, L.~Scorzato, C.~Urbach, and M.~Wagner [ETM Collaboration],
  JHEP {\bf 1304}, 137 (2013).

%
 \bibitem{a0}
A.~Abdel-Rehim, C.~Alexandrou, J.~Berlin, M.~Dalla Brida, M.~Gravina, M.~Wagner, 
 in {\it 32nd International Symposium on Lattice Field Theory, New York, 2014} 
 [arXiv:1410.8757[hep-lat]].

\bibitem{Sasa10} 
  S.~Prelovsek, T.~Draper, C.~B.~Lang, M.~Limmer, K.-F.~Liu, N.~Mathur, and D.~Mohler,
  Phys.\ Rev.\ D {\bf 82}, 094507 (2010).
  
 \bibitem{Chen06}
  Y.~Chen {\it et al.},
  Phys.\ Rev.\ D {\bf 73}, 014516 (2006).

\bibitem{Alexandrou:2006cq} 
  C.~Alexandrou, P.~de Forcrand and B.~Lucini,
  Phys.\ Rev.\ Lett.\  {\bf 97}, 222002 (2006)
  [hep-lat/0609004].

\bibitem{Jaffe:2004ph} 
  R.~L.~Jaffe,
  Phys.\ Rept.\  {\bf 409}, 1 (2005)
  [hep-ph/0409065].

\bibitem{Wagner:2011fs} 
  M.~Wagner and C.~Wiese [European Collaboration],
  JHEP {\bf 1107}, 016 (2011)
  [arXiv:1104.4921 [hep-lat]].
 
\bibitem{large Nc}
  Feng-K.~Guo, L.~Liu, Ulf-G.~Mei{\ss}ner and P.~Wang,
  Phys.\ Rev.\ D {\bf 88}, 074506 (2013).
  
\bibitem{Ali Khan:2000iz} 
  A.~Ali Khan {\it et al.}  [CP-PACS Collaboration],
  Phys.\ Rev.\ D {\bf 63}, 034502 (2000)
  [hep-lat/0008011].
  
\bibitem{dilution}
  J.~Foley, K.~J.~Juge, A.~$\acute{\rm O}$.~Cais, M.~Peardon, S.~M.~Ryan, and J.-I.~Skullerud  [TrinLat Collaboration],
  Comp.\ Phys.\ Comm. {\bf 172}, 145 (2005).

\bibitem{Luscher}
 M.~L\"ucher, Commun. Math. Phys. {\bf 104}, 177 (1986);  Commun.\  Math.\  Phys.\  {\bf 105}, 153 (1986);
Nucl.\  Phys.\   B {\bf  354}, 531  (1991).

\bibitem{Sharpe:1992pp} 
  S.~R.~Sharpe, R.~Gupta and G.~W.~Kilcup,
  Nucl.\ Phys.\ B {\bf 383}, 309 (1992).





  


\end{thebibliography}

\end{document}